# Universal scaling between precursory duration and event size across mechanically driven geohazards

Qinghua Lei[1], Didier Sornette[2]

[1]Department of Earth Sciences, Uppsala University, Uppsala, 752 36, Sweden
[2]Institute of Risk Analysis, Prediction and Management, Academy for Advanced Interdisciplinary Studies, Southern University of Science and Technology, Shenzhen, 518055, China

*Correspondence to*: Qinghua Lei (qinghua.lei@geo.uu.se)

**Abstract.** Many catastrophic events, including landslides, rockbursts, glacier breakoffs, and volcanic eruptions, are preceded by an observable acceleration phase that offers a critical window for early warning and hazard mitigation; however, the duration of this precursory phase remains poorly constrained across sites, scales, and hazard types. This limitation arises because the onset of acceleration is often identified using heuristic thresholds or empirical criteria. Here, we introduce a physics-based framework that objectively constrains the precursory duration from accelerating dynamics, without prescribing the onset a priori or being tied to any specific observable. We analyze a global dataset of 109 geohazard events across seven continents over the past century, quantifying their precursory durations in a consistent manner. For mechanically driven instabilities, we identify a robust scaling between precursory duration and failure volume spanning more than ten orders of magnitude. When expressed in terms of a characteristic system size, this relationship is close to linear, consistent with finite-size scaling near a dynamical critical point. This behavior indicates that precursory duration reflects the progressive growth of correlated deformation up to system-spanning scales, rather than local rupture kinetics. The resulting universality points to common organizing mechanisms governing the approach to catastrophic failure across mechanically driven geohazards.

## 1 Introduction

Forecasting catastrophic events in Earth systems remains challenging despite the widespread availability of high-resolution monitoring data. Across a wide range of geohazards, including landslides, rockbursts, glacier breakoffs, and volcanic eruptions, catastrophic failure is often preceded by a phase of accelerating deformation, seismicity, or other observable precursory signals (Lacroix et al., 2020; Feng et al., 2017; Acocella et al., 2023; Lei et al., 2023; Lei and Sornette, 2025a; Faillettaz et al., 2015). This precursory phase represents a critical window for early warning, emergency planning, and hazard mitigation. A fundamental yet largely unanswered question is the duration of this phase prior to final failure. Addressing this question is essential for understanding the temporal organization of failure processes and for constraining the limits of actionable lead time across geohazards.

Despite its central importance, the definition of a precursory phase and its duration remains poorly constrained. In many previous studies, the onset of a precursory phase is often identified using heuristic thresholds, visual inspection, or empirical criteria tied to specific observables (Faillettaz et al., 2015; Passarelli and Brodsky, 2012; Phillipson et al., 2013; Xue et al., 2014; Dick et al., 2015; Crosta et al., 2017; Krøgli et al., 2018; Askaripour et al., 2022; Valletta et al., 2022; Leinauer et al., 2023; Feng et al., 2025), rendering the inferred precursory duration subjective and difficult to compare across events. Different choices of observables, smoothing procedures, or reference times can lead to markedly different estimates of when precursory acceleration begins, thereby obscuring any systematic temporal organization of failure processes. As a result, this ambiguity has so far limited efforts to establish general constraints on precursory duration and to assess whether comparable preparatory processes exist across geohazards.

The precursory dynamics of a heterogeneous system are typically characterized by strong nonlinearity and intermittency, reflecting the presence of positive feedbacks and hierarchically organized damage accumulation as the system approaches failure (Johansen and Sornette, 2000; Sornette, 2002). These features have been successfully captured by the log-periodic power law singularity (LPPLS) model—a physics-based framework that simultaneously describes both the overall accelerating trend and the superimposed log-periodic organization of deformation, reflecting a discrete hierarchy of accelerating instabilities and temporary consolidation as rupture is approached in many geophysical systems (Sornette and Sammis, 1995; Ouillon and Sornette, 2000; Faillettaz et al., 2008; Lei and Sornette, 2025c, b, d). The LPPLS model has also been validated using a global dataset of 109 geohazard events (Lei and Sornette, 2025e), encompassing a wide range of hazard types (rockfalls, rockslides, soilslides, hanging glaciers, ice shelves, and volcanic eruptions), observable quantities (displacement, angular change, earthquake count, energy release, rift length, and gas emission), and monitoring approaches (ground-based measurements, remote sensing observations, and seismic/geochemical networks). This validation demonstrates its capability to describe accelerating dynamics across different site settings, spatiotemporal scales, observational resolutions, and hazard types.

Here, we build on this validated framework and dataset to quantify the precursory duration across diverse geohazards. Crucially, we leverage the LPPLS framework to render the concept of precursory duration explicit and operational by endogenising the identification of acceleration onset within the model calibration to monitoring data, rather than prescribing it a priori. This enables us to examine how precursory duration scales with event size, revealing a common scaling relationship that links characteristic time and length scales in mechanically driven instabilities. This study focuses on near-surface geohazards, including landslides, rockbursts, glacier breakoffs and volcanic eruptions, for which precursory acceleration is more clearly measurable and physically interpretable owing to shallow sources, spatially constrained rupture zones and dense monitoring. Earthquakes are not considered here, as the generality and interpretation of accelerating seismic release remain debated (Mignan, 2011).

## 2 Methodology

By explicitly capturing both the underlying accelerating trend and the superimposed oscillatory dynamics that characterize a geophysical crisis, the LPPLS model provides a physically grounded description of how instability develops over time as the system approaches catastrophic failure. The first-order expression of the LPPLS formulation is given by (Sornette and Sammis, 1995; Lei and Sornette, 2025e):

$$\Omega(t) = A + \left\{ B + C\cos\left[\omega \ln(t_c - t) - \phi\right] \right\} (t_c - t)^m \text{, with } m < 1 . \tag{1}$$

where $\Omega$ denotes the observable quantity (e.g., displacement, strain, energy release, earthquake count, or gas emission, etc.), $t$ is time, $t_c$ is the critical time, $m$ is the critical exponent, $\omega$ is the angular log-periodic frequency, $\phi$ is a phase shift, and $A$, $B$, and $C$ are constants. Thus, the LPPLS model involves seven parameters $\boldsymbol{\theta} = \{t_c, m, \omega, \phi, A, B, C\}$. It captures power law acceleration toward a finite-time singularity, modulated by log-periodic oscillations associated with discrete scale invariance and intermittent rupture processes in heterogeneous materials (Lei and Sornette, 2025c, b, d, e; Sornette, 1998). The oscillatory term represents a partial breaking of continuous scale invariance into discrete scale invariance (Sornette, 1998), consistent with rupture processes governed by hierarchical damage accumulation in heterogeneous materials.

To determine precursory duration, we calibrate the LPPLS model (Text S1) to time series of observations $\Omega(t_i)$ recorded at times $t_i \in [t_0, t_{end}]$, with $i = 1, \ldots, N$, where $N$ is the total number of time stamps, and $t_0$ and $t_{end}$ respectively denote the start and end times of the observational record. Importantly, the early portion of the time series may reflect dynamics unrelated to the precursory activity and therefore does not necessarily fall within the accelerating regime described by the LPPLS model. To objectively identify the onset of the acceleration, we adopt the Lagrange regularization approach (Demos and Sornette, 2017, 2019), where the starting time $\tau$ is treated as an endogenous variable and determined by minimizing a regularized cost function:

$$\chi^2(\boldsymbol{\theta}, \tau) = \frac{1}{n(\tau) - p} r^2(\boldsymbol{\theta}, \tau) - \lambda n(\tau) , \tag{2}$$

where $p = 7$ is the number of LPPLS model parameters, $n$ is the number of observations within the window $[\tau, t_{end}]$, $r^2$ is the sum of squared residuals of the LPPLS fit over this window, and $\lambda$ is a Lagrange multiplier controlling the trade-off between goodness of fit and window length. For each candidate $\tau$, the LPPLS model is calibrated over the interval $[\tau, t_{end}]$ using a stable and robust fitting scheme (Filimonov and Sornette, 2013), with the minimized $r^2$ further substituted into Eq. (2) to evaluate the regularized cost function. In our implementation, the Lagrange multiplier $\lambda$ is estimated as the slope in a linear regression of $r^2$ with respect to $\tau$. We scan $\tau$ over the interval from $t_0$ up to the latest time for which at least 30 data points remain available for calibration. The optimal value $\tau^*$, which minimizes $\chi^2$, is identified as the onset time, and the precursory duration is then estimated as $T = t_{end} - \tau^*$. For presentation purposes, we compute a normalized regularized cost function $\tilde{\chi}^2$, by rescaling $\chi^2$ using its minimum and maximum values over the scanned range of $\tau$; this normalization does not affect the identification of $\tau^*$.

The Lagrange regularization approach admits a natural interpretation from the perspective of statistical physics (Gibbs, 1902; Callen, 1985). Specifically, fixing the calibration window length is analogous to working in a canonical ensemble with a fixed number of degrees of freedom, whereas allowing the window start time to vary corresponds to a grand canonical ensemble in which the number of data points is free but incurs an effective cost controlled by the Lagrange multiplier $\lambda$. In this analogy, $\lambda$ plays a role similar to a chemical potential that counteracts the intrinsic bias toward short windows. Shorter windows yield better fits, not because they are more informative, but because a fixed number of parameters is calibrated on fewer data points. The Lagrange multiplier term $\lambda$ therefore penalizes excessively short windows, preventing a systematic bias toward them and enabling an objective identification of the onset of the precursory phase without imposing it a priori.

## 3 Results

### 3.1 Objective identification of precursory duration across geohazards

We apply the LPPLS framework to four well-instrumented case studies spanning contrasting hazard types, illustrating how precursory duration can be objectively identified from monitored accelerating dynamics. The same analysis is then extended to other cases in the global dataset.

The first case study concerns the Veslemannen landslide in western Norway, a well-instrumented slope instability developed in high-grade metamorphic rocks. This site has been continuously monitored since 2014 using ground-based interferometric synthetic-aperture radar (Kristensen et al., 2021), providing displacement measurements with an accuracy of ~0.5 mm. In the months leading up to the collapse on 5 September 2019 (failure volume ~54,000 $m^3$), the landslide exhibited pronounced acceleration punctuated by intermittent deceleration episodes, resulting in cumulative displacements of several metres over a period of ~3 months (Fig. 1a shows data from one representative radar point, while Fig. S1 shows all other radar points). Applying the LPPLS model to the displacement time series captures both the accelerating and oscillatory dynamics characteristic of the precursory phase. Scanning the calibration window reveals a distinct minimum in the regularized cost function (Fig. 1), providing an estimate of the onset time of precurosry accelartion on 4 June 2019. The same analysis is performed for all other radar points (Fig. S1), with the earliest one (May 19, 2019) selected as the onset time for this event. This selected onset time is consistent with the convergence of onset estimates from all radar points that exhibit a stable plateau after August 2019 (Fig. S2a), demonstrating the robustness of the identified precursory onset. The interval between the onset and failure times then defines the precursory duration, which is 109 days.

The second case study examines a rockburst event that occurred in an underground coal mine in New South Wales, Australia. This mine, located at a depth of ~250 m and excavated using the longwall mining method, was instrumented with multipoint extensometers (with an accuracy of ~0.5 mm) to continuously monitor roof displacement within a gateroad (Shen et al., 2008). In the days preceding the roof collapse at 13:35 on 4 June 2004, the displacement time series exhibited a clear acceleration punctuated by intermittent deceleration episodes (Fig. 1b), indicative of an evolving instability. The LPPLS model captures both the acceleration and oscillation patterns characteristic of the precursory phase. A pronounced minimum in the regularized cost function indicates an onset of acceleration at 2:00 on 2 June 2004 (Fig. 1b), supported by the convergence of onset estimates toward stable plateaus (Fig. S2b), implying a precursory duration of ~2.5 days. The associated failure volume is estimated to be ~50 m$^3$, based on the dimensions of the gateroad (5.2 m in width and 3 m in height).

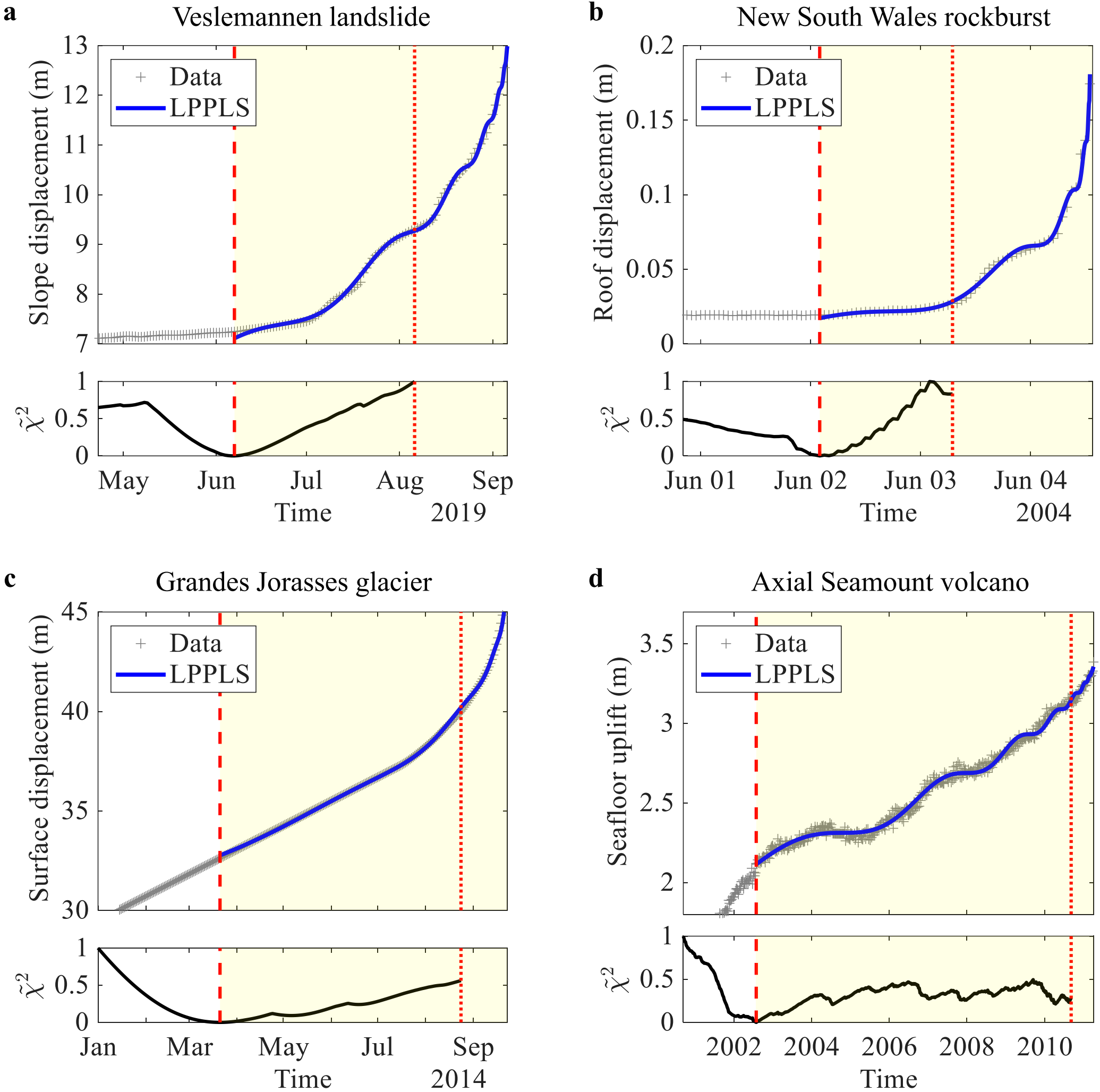

**Figure 1: Representative examples of precursory dynamics and onset identification across geohazards. In each subfigure, the upper panel shows the monitoring data (grey symbols) together with the corresponding LPPLS model fit (blue line), while the lower panel shows the normalized regularized cost function $\tilde{\chi}^2$ as a function of the calibration window start time $\tau$. Vertical dashed lines indicate the identified onset time $\tau^*$ of precursory acceleration, vertical dotted lines mark the latest time of the $\tau$-scanning interval, and shaded regions highlight the inferred precursory phase. The examples include: (a) slope surface displacement (daily aggregation) prior to the catastrophic landslide on 5 September 2019, at Veslemannen, Norway; (b) tunnel roof displacement (hourly aggregation) prior to a violent rockburst on 4 June 2004, in an underground coal mine in New South Wales, Australia; (c) glacier surface displacement (daily aggregation) prior to rapid breakoffs in late September 2014, at the Grandes Jorasses glacier, Italy; and (d) seafloor uplift (weekly aggregation) prior to the volcanic eruption on 6 April 2011, at Axial Seamount, Pacific Ocean.**

The third case study regards a hanging cold glacier located on the south face of the Grandes Jorasses massif, Italy, at an elevation of ~3,950 m above sea level (Faillettaz et al., 2015). Surface displacements were monitored using a robotic total station equipped with multiple reflectors, providing measurements with an accuracy of ~1 cm. In 2014, a large ice volume of ~$1.05\times10^5$ $m^3$ detached in two consecutive breakoff events (Faillettaz et al., 2016) on September 23 and 29. In the months preceding the final event, the glacier exhibited a smooth, sustained acceleration in surface displacement that is well captured by the LPPLS model (Fig. 1c). A clear minimum in the regularized cost function identifies the onset of acceleration in late March 2014, corresponding to a precursory duration of ~6 months. The inferred onset time exhibits a stable plateau from late August 2014 onward, indicating robust identification of the precursory onset (Fig. S2c).

The fourth case study focuses on Axial Seamount, an active submarine volcano located ~500 km offshore Oregon, USA, whose summit caldera lies at a water depth of about 1.5 km. Axial Seamount has been continuously monitored since 1998 by a cabled seafloor observatory, which records vertical seafloor deformation with a resolution of ~1 cm using bottom pressure recorders (Chadwick et al., 2022). We focus on the eruption that occurred in April 2011, prior to which the volcano exhibited a prolonged sequence of inflation–deflation cycles spanning more than eight years (Chadwick et al., 2012) (Fig. 1d). The erupted magma volume associated with the 2011 event is ~$9.9\times10^7$ $m^3$. The seafloor uplift time series is well described by the LPPLS model, capturing the long-term acceleration together with the superimposed oscillations characteristic of episodic magma pressurisation and release during volcanic unrest. A well-defined minimum in the regularized cost function identifies the onset of precursory acceleration in 2002, yielding a precursory duration of ~8.7 years. The inferred onset time is already detectable prior to 2007 and, despite fluctuations between 2007 and 2010, stabilises thereafter, indicating robust identification of the precursory onset (Fig. S2d).

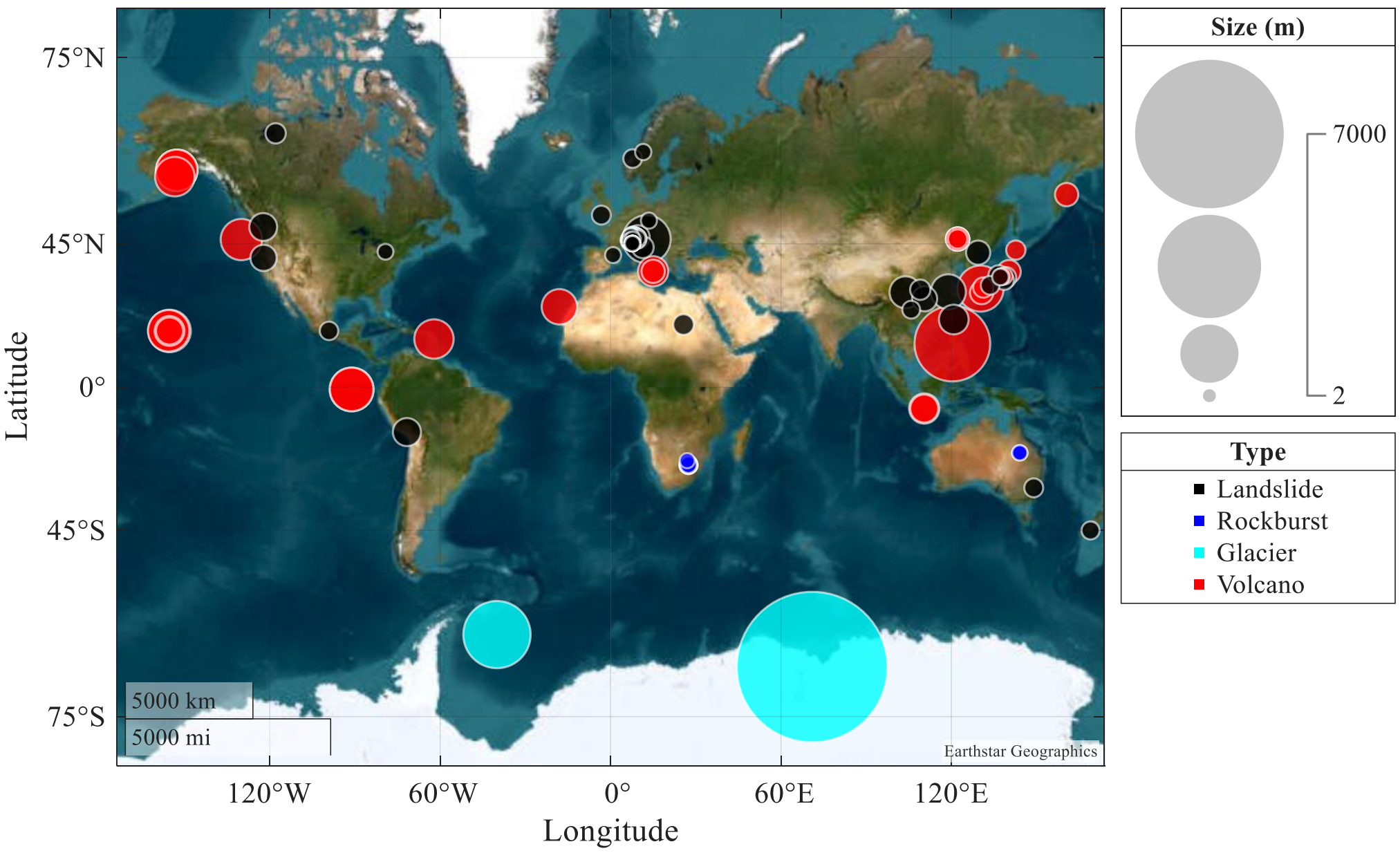

**Figure 2: Global distribution of 109 historical geohazard events. The dataset includes 49 landslides, 11 rockbursts, 17 glacier breakoffs, and 32 volcanic eruptions. Colors indicate hazard type and symbol size scales with the cube root of failure volume.**

The four case studies presented above form part of a larger global dataset comprising 109 geohazard events, including landslides, rockbursts, glacier breakoffs, and volcanic eruptions, distributed across seven continents (Fig. 2). The same LPPLS-based framework is applied consistently to all events in the dataset to identify their onset times and precursory durations. Details of the full dataset, including event locations, hazard types, data sources, failure volumes, and precursory durations, are provided in the supplement (Text S2 and Tables S1–S4).

### 3.2 Scaling between precursory duration and failure volume

We perform a correlation analysis between precursory duration $T$ and failure volume $V$ across the global dataset of 109 geohazard events. When volcanic eruptions are excluded, the remaining mechanically driven instabilities, including landslides, rockbursts, and glaciers, exhibit a systematic scaling between precursory duration and failure volume (Fig. 3). Despite spanning more than ten orders of magnitude in failure volume and encompassing diverse hazard types, the data collapse onto a clear scaling trend (Fig. 3). A log-log regression yields a power law relation $T \propto V^{\zeta}$, with $\zeta = 0.35\pm0.05$ (Fig. 3). The statistical significance of the observed scaling is supported by multiple complementary statistical tests (Text S3 and Table S5), all of which decisively reject the null hypothesis of no dependence between $T$ and $V$.

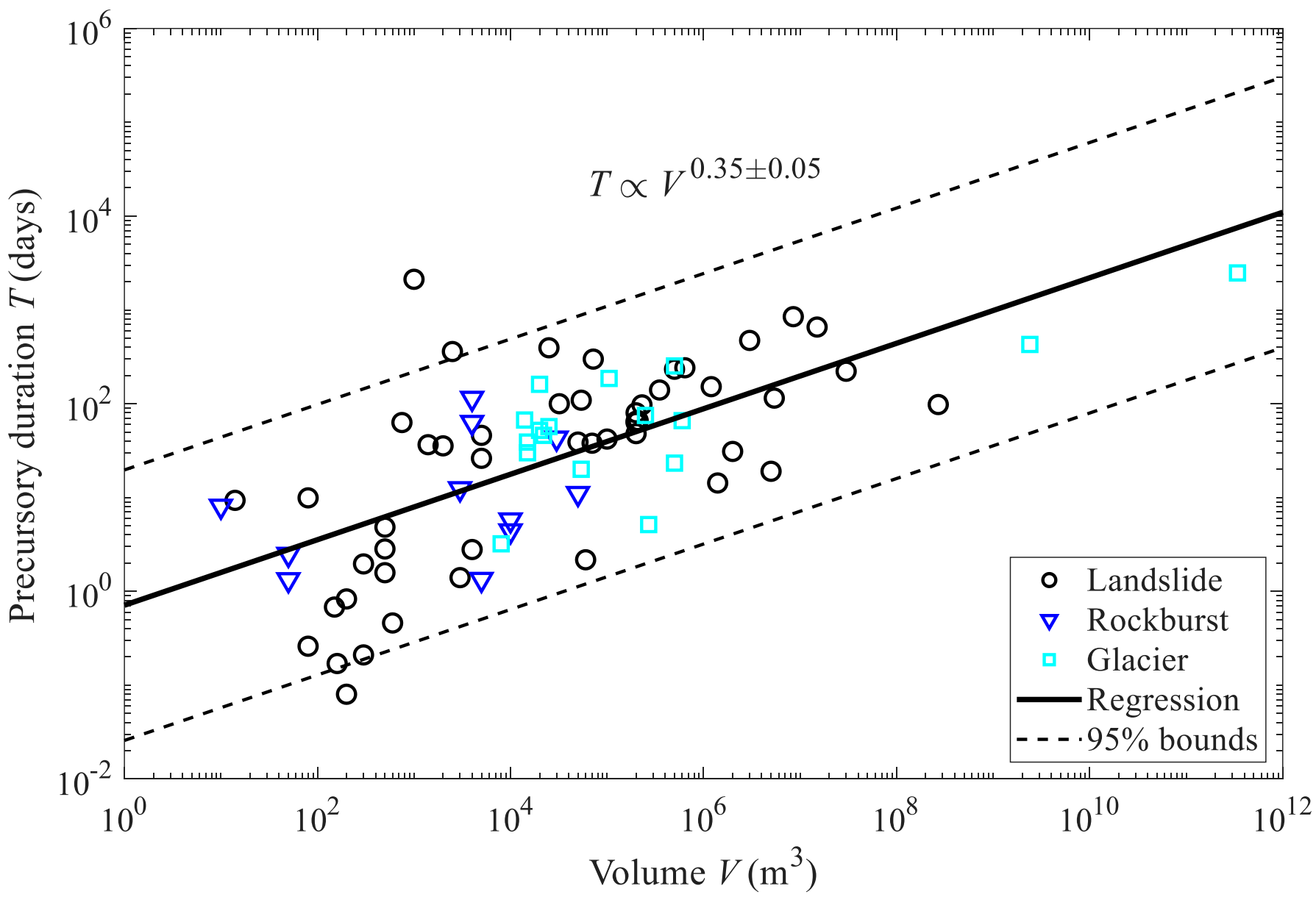

**Figure 3: Scaling relationship between precursory duration $T$ and failure volume $V$ for mechanically driven geohazards. Symbols indicate hazard type, the solid black line shows the best-fit power law regression $T \propto V^{0.35\pm0.05}$, and dashed lines mark the 95% confidence interval.**

When analysed by hazard type (Table S5), landslides, glaciers, and rockbursts yield scaling exponents that cluster within $\zeta = 0.30\pm0.10$, broadly consistent with the exponent obtained for the combined dataset of mechanically driven instabilities, although the rockburst estimate remains statistically inconclusive, likely owing to limited sample size. Volcanic eruptions, by contrast, do not exhibit a comparable dependence of precursory duration on failure volume: their fitted exponent is close to zero, and the null hypothesis of no scaling cannot be rejected (Table S5). Notably, although volcanic eruptions do not exhibit a significant scaling, their precursory durations largely fall within the confidence bounds of the scaling derived from mechanically driven instabilities (Fig. S3), indicating compatibility with the overall trend despite the lack of a clear correlation. Taken together, these results demonstrate a statistically robust scaling between precursory duration and failure volume for mechanically driven geohazards, while delineating volcanic eruptions as a distinct class in which failure volume is not the primary control on precursory timescales.

**4 Discussion**

Our analysis reveals a clear empirical scaling between precursory duration and failure volume across mechanically driven geohazards, $T \propto V^{0.35}$, which implies $T \propto L^{1.05}$ when expressing volume in terms of a characteristic system size $L \sim V^{1/3}$. This scaling behaviour implies an "effective" preparatory rate, defined as $\gamma^* = L/T$, characterizing the pace at which instability progressively organizes across a finite system. From the observed scaling, we estimate $\gamma^*$ to be ~0.85 m/day, with a 95% confidence interval spanning approximately 0.03 to 23 m/day. This variability is expected and reflects differences in system geometry, material properties, and loading conditions. Notably, this scaling emerges only when precursory duration is defined consistently within a unified, physics-based framework, rather than through heuristic or subjective criteria commonly used in previous studies (Faillettaz et al., 2015; Passarelli and Brodsky, 2012; Phillipson et al., 2013; Xue et al., 2014; Dick et al., 2015; Crosta et al., 2017; Askaripour et al., 2022; Valletta et al., 2022; Leinauer et al., 2023; Feng et al., 2025).

The observed near-linear $L$–$T$ scaling admits a natural interpretation within the framework of dynamical critical phenomena and finite-size scaling. Catastrophic failure can be viewed as a critical point marking the transition from a quasi-static creep regime to an elasto-dynamic rupture regime, culminating in system-wide collapse (Bonamy and Bouchaud, 2011; Johansen and Sornette, 2000; Rundle et al., 2003; Sornette, 2002; Vu et al., 2019). As a system approaches the critical rupture time $t_c$, it is characterized by the progressive development of correlated clusters of damage and/or slip, quantified by a correlation length $\xi$ and an associated correlation time $\mathcal{T}$, which sets the characteristic timescale of fluctuations and precursory activity (Alava et al., 2006; Girard et al., 2010; Pradhan et al., 2010; Sornette, 2006; Sornette and Andersen, 1998). For idealized infinite systems, these two quantities diverge as power laws of the distance to criticality (Cardy, 1996; Henkel et al., 2008; Privman, 1990), $\xi(t) \sim (t_c - t)^{-\nu}$ and $\mathcal{T}(t) \sim (t_c - t)^{-\mu}$, where $\nu$ and $\mu$ are the critical exponents governing spatial and temporal correlations, respectively. These divergences can be compactly summarized by the dynamic scaling relation $\mathcal{T} \sim \xi^z$, where $z = \mu/\nu$ is the dynamic critical exponent. However, real geophysical systems are finite. Finite-size scaling theory predicts that the divergence of $\xi$ cannot continue indefinitely: as $t \to t_c$, $\xi$ saturates at the characteristic system size, i.e.

$\xi \lesssim L$ . By the same reasoning, the correlation time also saturates at a finite value controlled by the system size, $\mathcal{T} \sim \xi^z \lesssim L^z$ , such that the maximal observable precursory duration is set by this finite-size cutoff.

Our empirically observed near-linear *L*–*T* relation therefore corresponds to a dynamic exponent $z \approx 1$, implying that temporal and spatial correlations grow approximately proportionally as instability organizes toward system-spanning rupture. In this finite-size criticality view, the precursory phase reflects the approach to a critical point whose apparent "duration" is determined by the time required for correlations to reach the system scale. This perspective contrasts with traditional views in which precursory duration is controlled by local fracturing or sliding rates, as emphasized in subcritical crack growth model (Brantut et al., 2013) and rate-and-state friction model (Marone, 1998), and instead highlights the dominant role of system-scale organization in setting precursory timescales. Such behaviour is consistent with rupture processes governed by long-range interactions, where stress perturbations are transmitted efficiently across the system (Kawamura et al., 2012; Sornette and Andersen, 1998). The growth of correlations is not diffusive but occurs through cascades of interacting local instabilities, leading to hierarchical organization and intermittent activity (Ben-Zion, 2008; Sornette, 2002). The LPPLS framework, which captures both accelerating and oscillatory dynamics, is consistent with this picture, reflecting the coexistence of positive feedback arising from amplified interactions across scales and hierarchically organized damage/slip processes that evolve intermittently as failure is approached (Lei and Sornette, 2025e; Sornette, 1998).

Under this interpretation, the near-linear scaling between precursory duration and system size reflects the time required for instability to organize at the system scale, rather than the rate of rupture propagation. Accordingly, precursory duration is primarily constrained by system size, even though the underlying deformation and rupture processes remain highly nonlinear and intermittent. This scaling is therefore expected for geohazards in which failure develops through progressive damage accumulation within a finite source region, including landslides, rockbursts, and glacier breakoffs. The absence of a significant scaling for volcanic eruptions is also physically informative, consistent with the fact that volcanic unrest is commonly controlled by magma transport, pressurization, phase changes, and evolving rheology (Caricchi et al., 2021), processes that do not necessarily involve a progressive, system-wide build-up of mechanically correlated deformation within a fixed source volume. Taken together, our results identify the scaling between precursory duration and system size as a robust signature of finite-size criticality in a broad class of mechanically driven instabilities.

The recognition of finite-size scaling and system-scale precursory dynamics has important implications for the interpretation of geohazard monitoring observations. In many geohazard contexts, particularly landslides and glacier breakoffs, it is commonly assumed that surface measurements provide only indirect or proxy information about subsurface failure processes, motivating efforts to directly access the presumed rupture plane through boreholes. Our results challenge this view. In mechanically driven instabilities, the rupture surface is not a pre-existing object that can be interrogated during the precursory phase, but rather an emergent feature that forms only at final failure. The observed scaling indicates that the precursory phase reflects the progressive growth of correlated deformation up to system-spanning scales, rather than localized rupture along a nascent failure plane. Consequently, provided they are located within the actively deforming zone, surface-based measurements—including extensometers, reflectors, GNSS stations, and radar interferometry—directly sample the relevant system-scale precursory dynamics. This provides a physical foundation for the long-standing empirical success of surface monitoring and forecasting approaches for landslides and glacier breakoffs (Crosta and Agliardi, 2003; Faillettaz et al., 2015, 2016; Loew et al., 2017), and suggests that the predictive value of such observations arises from critical correlation buildup across the entire unstable mass, rather than from their proximity to an elusive failure plane.

## Code and data availability

No new data were produced in this work. All data underlying this study were obtained from previously published sources, with full details and references provided in the supplementary material. The code used in this study is available at Zenodo https://doi.org/10.5281/zenodo.15362582.

## Supplement link

The link to the supplement will be included by Copernicus.

## Author contributions

QL conceived the study, performed the analysis, and wrote the manuscript. QL and DS devised the method and interpreted the results. DS reviewed and edited the manuscript.

## Competing interests

The authors declare that they have no conflict of interest.

## Disclaimer

**Acknowledgement**

During manuscript preparation, the authors used ChatGPT to assist with language editing and readability. The authors reviewed and edited the manuscript and take full responsibility for its content.

**Financial support**

This work was supported by the European Research Council (ERC Consolidator Grant, grant no. 101232311) and the Norwegian Water Resources and Energy Directorate. The research was also partially supported by the National Natural Science Foundation of China (Grant No. U2039202, T2350710802) and the Shenzhen Science and Technology Innovation Commission (Grant No. GJHZ20210705141805017).

**References**

Acocella, V., Ripepe, M., Rivalta, E., Peltier, A., Galetto, F., and Joseph, E.: Towards scientific forecasting of magmatic eruptions, Nat. Rev. Earth. Environ., 5, 5–22, https://doi.org/10.1038/s43017-023-00492-z, 2023.

Alava, M. J., Nukala, P. K. V. V., and Zapperi, S.: Statistical models of fracture, Advances in Physics, 55, 349–476, https://doi.org/10.1080/00018730300741518, 2006.

Askaripour, M., Saeidi, A., Rouleau, A., and Mercier-Langevin, P.: Rockburst in underground excavations: A review of mechanism, classification, and prediction methods, Undergr. Space, 7, 577–607, https://doi.org/10.1016/j.undsp.2021.11.008, 2022.

Ben-Zion, Y.: Collective behavior of earthquakes and faults: Continuum-discrete transitions, progressive evolutionary changes, and different dynamic regimes, Reviews of Geophysics, 46, 2008RG000260, https://doi.org/10.1029/2008RG000260, 2008.

Bonamy, D. and Bouchaud, E.: Failure of heterogeneous materials: A dynamic phase transition?, Physics Reports, 498, 1–44, https://doi.org/10.1016/j.physrep.2010.07.006, 2011.

Brantut, N., Heap, M. J., Meredith, P. G., and Baud, P.: Time-dependent cracking and brittle creep in crustal rocks: A review, J Struct. Geol., 52, 17–43, https://doi.org/10.1016/j.jsg.2013.03.007, 2013.

Callen, H. B.: Thermodynamics and an Introduction to Thermostatistics, John Wiley & Sons, New York, 1985.

Cardy, J.: Scaling and Renormalization in Statistical Physics, 1st ed., Cambridge University Press, Cambridge, https://doi.org/10.1017/CBO9781316036440, 1996.

Caricchi, L., Townsend, M., Rivalta, E., and Namiki, A.: The build-up and triggers of volcanic eruptions, Nat. Rev. Earth. Environ., 2, 458–476, https://doi.org/10.1038/s43017-021-00174-8, 2021.

Chadwick, W. W., Nooner, S. L., Butterfield, D. A., and Lilley, M. D.: Seafloor deformation and forecasts of the April 2011 eruption at Axial Seamount, Nat. Geosci., 5, 474–477, https://doi.org/10.1038/ngeo1464, 2012.

Chadwick, W. W., Wilcock, W. S. D., Nooner, S. L., Beeson, J. W., Sawyer, A. M., and Lau, T. -K.: Geodetic monitoring at Axial Seamount since its 2015 eruption reveals a waning magma supply and tightly linked rates of deformation and seismicity, Geochem. Geophys. Geosyst., 23, e2021GC010153, https://doi.org/10.1029/2021GC010153, 2022.

Crosta, G. B. and Agliardi, F.: Failure forecast for large rock slides by surface displacement measurements, Can. Geotech. J., 40, 176–191, https://doi.org/10.1139/t02-085, 2003.

Crosta, G. B., Agliardi, F., Rivolta, C., Alberti, S., and Dei Cas, L.: Long-term evolution and early warning strategies for complex rockslides by real-time monitoring, Landslides, 14, 1615–1632, https://doi.org/10.1007/s10346-017-0817-8, 2017.

Demos, G. and Sornette, D.: Birth or burst of financial bubbles: which one is easier to diagnose?, Quantitative Finance, 17, 657–675, https://doi.org/10.1080/14697688.2016.1231417, 2017.

Demos, G. and Sornette, D.: Comparing nested data sets and objectively determining financial bubbles' inceptions, Physica A Stat. Mech. Appl., 524, 661–675, https://doi.org/10.1016/j.physa.2019.04.050, 2019.

Dick, G. J., Eberhardt, E., Cabrejo-Liévano, A. G., Stead, D., and Rose, N. D.: Development of an early-warning time-of-failure analysis methodology for open-pit mine slopes utilizing ground-based slope stability radar monitoring data, Can. Geotech. J., 52, 515–529, https://doi.org/10.1139/cgj-2014-0028, 2015.

Faillettaz, J., Pralong, A., Funk, M., and Deichmann, N.: Evidence of log-periodic oscillations and increasing icequake activity during the breaking-off of large ice masses, J. Glaciol., 54, 725–737, https://doi.org/10.3189/002214308786570845, 2008.

Faillettaz, J., Funk, M., and Vincent, C.: Avalanching glacier instabilities: Review on processes and early warning perspectives, Rev. Geophys., 53, 203–224, https://doi.org/10.1002/2014RG000466, 2015.

Faillettaz, J., Funk, M., and Vagliasindi, M.: Time forecast of a break-off event from a hanging glacier, Cryosphere, 10, 1191–1200, https://doi.org/10.5194/tc-10-1191-2016, 2016.

Feng, B., Zeng, P., Li, T., Sun, X., Zhu, X., Fan, X., and Xu, Q.: Acceleration stage detection and dynamic model selection for real-time landslide time-of-failure predictions, Acta Geotech., 2025.

Feng, X.-T., Liu, J., Chen, B., Xiao, Y., Feng, G., and Zhang, F.: Monitoring, warning, and control of rockburst in deep metal mines, Engineering, 3, 538–545, https://doi.org/10.1016/J.ENG.2017.04.013, 2017.

Filimonov, V. and Sornette, D.: A stable and robust calibration scheme of the log-periodic power law model, Physica A Stat. Mech. Appl., 392, 3698–3707, https://doi.org/10.1016/j.physa.2013.04.012, 2013.

Gibbs, J. W.: Elementary Principles in Statistical Mechanics, Charles Scribner's Sons, New York, 1902.

Girard, L., Amitrano, D., and Weiss, J.: Failure as a critical phenomenon in a progressive damage model, J. Stat. Mech., 2010, P01013, https://doi.org/10.1088/1742-5468/2010/01/P01013, 2010.

Henkel, M., Hinrichsen, H., and Lübeck, S.: Non-Equilibrium Phase Transitions, Springer Netherlands, Dordrecht, https://doi.org/10.1007/978-1-4020-8765-3, 2008.

Johansen, A. and Sornette, D.: Critical ruptures, Eur. Phys. J. B, 18, 163–181, https://doi.org/10.1007/s100510070089, 2000.

Kawamura, H., Hatano, T., Kato, N., Biswas, S., and Chakrabarti, B. K.: Statistical physics of fracture, friction, and earthquakes, Rev. Mod. Phys., 84, 839–884, https://doi.org/10.1103/RevModPhys.84.839, 2012.

Kristensen, L., Czekirda, J., Penna, I., Etzelmüller, B., Nicolet, P., Pullarello, J. S., Blikra, L. H., Skrede, I., Oldani, S., and Abellan, A.: Movements, failure and climatic control of the Veslemannen rockslide, Western Norway, Landslides, 18, 1963–1980, https://doi.org/10.1007/s10346-020-01609-x, 2021.

Krøgli, I. K., Devoli, G., Colleuille, H., Boje, S., Sund, M., and Engen, I. K.: The Norwegian forecasting and warning service for rainfall- and snowmelt-induced landslides, Nat. Hazards Earth Syst. Sci., 18, 1427–1450, https://doi.org/10.5194/nhess-18-1427-2018, 2018.

Lacroix, P., Handwerger, A. L., and Bièvre, G.: Life and death of slow-moving landslides, Nat. Rev. Earth Environ., 1, 404–419, https://doi.org/10.1038/s43017-020-0072-8, 2020.

Lei, Q. and Sornette, D.: Endo-exo framework for a unifying classification of episodic landslide movements: Implications for forecasting catastrophic failures, Sci. Adv., https://doi.org/10.1126/sciadv.ady9141, 2025a.

Lei, Q. and Sornette, D.: Log-periodic power law singularities in landslide dynamics: Statistical evidence from 52 crises, Geophysical Research Letters, 52, e2025GL116379, https://doi.org/10.1029/2025GL116379, 2025b.

Lei, Q. and Sornette, D.: Log-periodic signatures prior to volcanic eruptions: evidence from 34 events, Earth and Planetary Science Letters, 666, 119496, https://doi.org/10.1016/j.epsl.2025.119496, 2025c.

Lei, Q. and Sornette, D.: Oscillatory finite-time singularities in rockbursts, Int. J. Rock Mech. Min. Sci., 192, 106156, https://doi.org/10.1016/j.ijrmms.2025.106156, 2025d.

Lei, Q. and Sornette, D.: Unified failure model for landslides, rockbursts, glaciers, and volcanoes., Commun. Earth Environ, 6, 390, https://doi.org/10.1038/s43247-025-02369-z, 2025e.

Lei, Q., Sornette, D., Yang, H., and Loew, S.: Real-time forecast of catastrophic landslides via dragon-king detection, Geophys. Res. Lett., 50, e2022GL100832, https://doi.org/10.1029/2022GL100832, 2023.

Leinauer, J., Weber, S., Cicoira, A., Beutel, J., and Krautblatter, M.: An approach for prospective forecasting of rock slope failure time, Commun. Earth Environ., 4, 253, https://doi.org/10.1038/s43247-023-00909-z, 2023.

Loew, S., Gschwind, S., Gischig, V., Keller-Signer, A., and Valenti, G.: Monitoring and early warning of the 2012 Preonzo catastrophic rockslope failure, Landslides, 14, 141–154, https://doi.org/10.1007/s10346-016-0701-y, 2017.

Marone, C.: Laboratory-derived friction laws and their application to seismic faulting, Annu. Rev. Earth Planet. Sci., 26, 643–696, https://doi.org/10.1146/annurev.earth.26.1.643, 1998.

Mignan, A.: Retrospective on the Accelerating Seismic Release (ASR) hypothesis: Controversy and new horizons, Tectonophysics, 505, 1–16, https://doi.org/10.1016/j.tecto.2011.03.010, 2011.

Ouillon, G. and Sornette, D.: The concept of “critical earthquakes” applied to mine rockbursts with time-to-failure analysis, Geophys. J. Int., 143, 454–468, https://doi.org/10.1046/j.1365-246X.2000.01257.x, 2000.

Passarelli, L. and Brodsky, E. E.: The correlation between run-up and repose times of volcanic eruptions, Geophysical Journal International, 188, 1025–1045, https://doi.org/10.1111/j.1365-246X.2011.05298.x, 2012.

Phillipson, G., Sobradelo, R., and Gottsmann, J.: Global volcanic unrest in the 21st century: An analysis of the first decade, Journal of Volcanology and Geothermal Research, 264, 183–196, https://doi.org/10.1016/j.jvolgeores.2013.08.004, 2013.

Pradhan, S., Hansen, A., and Chakrabarti, B. K.: Failure processes in elastic fiber bundles, Rev. Mod. Phys., 82, 499–555, https://doi.org/10.1103/RevModPhys.82.499, 2010.

Privman, V.: Finite Size Scaling and Numerical Simulation of Statistical Systems, World Scientific Publishing, Singapore, https://doi.org/10.1142/1011, 1990.

Rundle, J. B., Turcotte, D. L., Shcherbakov, R., Klein, W., and Sammis, C.: Statistical physics approach to understanding the multiscale dynamics of earthquake fault systems, Rev. Geophys., 41, 2003RG000135, https://doi.org/10.1029/2003RG000135, 2003.

Shen, B., King, A., and Guo, H.: Displacement, stress and seismicity in roadway roofs during mining-induced failure, Int. J. Rock Mech. Min. Sci., 45, 672–688, https://doi.org/10.1016/j.ijrmms.2007.08.011, 2008.

Sornette, D.: Discrete-scale invariance and complex dimensions, Phys. Rep., 297, 239–270, https://doi.org/10.1016/S0370-1573(97)00076-8, 1998.

Sornette, D.: Predictability of catastrophic events: Material rupture, earthquakes, turbulence, financial crashes, and human birth, Proc. Natl. Acad. Sci., 99, 2522–2529, https://doi.org/10.1073/pnas.022581999, 2002.

Sornette, D.: Critical Phenomena in Natural Sciences - Chaos, Fractals, Selforganization and Disorder: Concepts and Tools, Springer, Berlin/Heidelberg, https://doi.org/10.1007/3-540-33182-4, 2006.

Sornette, D. and Andersen, J. V.: Scaling with respect to disorder in time-to-failure, Eur. Phys. J. B, 1, 353–357, https://doi.org/10.1007/s100510050194, 1998.

Sornette, D. and Sammis, C. G.: Complex critical exponents from renormalization group theory of earthquakes: Implications for earthquake predictions, J. Phys. I France, 5, 607–619, https://doi.org/10.1051/jp1:1995154, 1995.

Valletta, A., Carri, A., and Segalini, A.: Definition and application of a multi-criteria algorithm to identify landslide acceleration phases, Georisk: Assessment and Management of Risk for Engineered Systems and Geohazards, 16, 555–569, https://doi.org/10.1080/17499518.2021.1952610, 2022.

Vu, C.-C., Amitrano, D., Plé, O., and Weiss, J.: Compressive failure as a critical transition: Experimental evidence and mapping onto the universality class of depinning, Phys. Rev. Lett., 122, 015502, https://doi.org/10.1103/PhysRevLett.122.015502, 2019.

Xue, L., Qin, S., Li, P., Li, G., Adewuyi Oyediran, I., and Pan, X.: New quantitative displacement criteria for slope deformation process: From the onset of the accelerating creep to brittle rupture and final failure, Engineering Geology, 182, 79–87, https://doi.org/10.1016/j.enggeo.2014.08.007, 2014.

Supplement of

# Universal scaling between precursory duration and event size across mechanically driven geohazards

Qinghua Lei[1], Didier Sornette[2]

[1]Department of Earth Sciences, Uppsala University, Uppsala, 752 36, Sweden
[2]Institute of Risk Analysis, Prediction and Management, Academy for Advanced Interdisciplinary Studies, Southern University of Science and Technology, Shenzhen, 518055, China

## Contents of this file

## Introduction

This document provides supplementary materials to complement the methods, results, and discussions in the main text. Text S1 details the LPPLS fitting procedure, Text S2 describes the dataset used in this study, and Text S3 presents the methodology for the statistical assessment of the scaling relation between precursory duration and failure volume. Figure S1 presents LPPLS calibration results and onset identification for the Veslemannen landslide (radar points 1, 3, 4, 5, 6, and 7). Figure S2 displays the temporal evolution of the inferred onset time for the Veslemannen landslide, the New South Wales rockburst, the Grandes Jorasses glacier, and the Axial Seamount volcano. Figure S3 compares volcanic precursory durations with scaling derived from mechanically driven geohazards. Tables S1–S4 summarize key information for the 109 geohazard events analyzed, including location, hazard type, material, volume, precursory duration, and references. Table S5 reports the results of the statistical tests of the scaling exponent for the full dataset including all hazard types and for each hazard type analyzed independently.

**Text S1.** Calibration scheme for the LPPLS model

Consider the time series data of observations $\Omega(t_i)$ recorded at times $t_i \in [\tau, t_{\text{end}}]$, with $i = 1, \ldots, n$, where $n$ is the number of time stamps, $\tau$ and $t_{\text{end}}$ respectively denote the start and end times of the time window for fitting the LPPLS model. We rewrite the LPPLS formula as:

$$\Omega(t) = A + B(t_c - t)^m + C_1(t_c - t)^m \cos\left[\omega \ln(t_c - t)\right] + C_2(t_c - t)^m \sin\left[\omega \ln(t_c - t)\right], \quad \text{(S1)}$$

where $C_1 = C\cos\phi$ and $C_2 = C\sin\phi$, such that the parameter set $\boldsymbol{\theta} = \{A, B, C, t_c, m, \omega, \phi\}$ (with three linear and four nonlinear parameters) of the original LPPLS formula given by equation (1) in the main text is transformed into a new parameter set $\boldsymbol{\theta} = \{A, B, C_1, C_2, t_c, m, \omega\}$ (with four linear and three nonlinear parameters). We define the sum of squared residuals as the cost function:

$$r^2(\boldsymbol{\theta}; \boldsymbol{\Omega}, \mathbf{t}) = \sum_{i=1}^{n} \left\{\Omega_i - A - B(t_c - t_i)^m - C_1(t_c - t_i)^m \cos\left[\omega \ln(t_c - t_i)\right] - C_2(t_c - t_i)^m \sin\left[\omega \ln(t_c - t_i)\right]\right\}^2, \quad \text{(S2)}$$

which is minimized using the ordinary least squares method to estimate the model parameters:

$$\hat{\boldsymbol{\theta}} = \underset{\boldsymbol{\theta}}{\arg\min}\, r^2(\boldsymbol{\theta}; \boldsymbol{\Omega}, \mathbf{t}). \quad \text{(S3)}$$

The four linear parameters $\{A, B, C_1, C_2\}$ are enslaved to the three nonlinear ones $\{t_c, m, \omega\}$, such that the minimization problem is reduced to:

$$\{\hat{t}_c, \hat{m}, \hat{\omega}\} = \underset{t_c, m, \omega}{\arg\min}\, \rho^2(t_c, m, \omega), \quad \text{(S4)}$$

where the profiled cost function is defined as:

$$\rho^2(t_c, m, \omega) = \min_{A, B, C_1, C_2} r^2(A, B, C_1, C_2, t_c, m, \omega) = r^2(t_c, m, \omega, \hat{A}, \hat{B}, \hat{C}_1, \hat{C}_2). \quad \text{(S5)}$$

The estimates for the four linear parameters can be obtained by solving the optimization problem for fixed values of the nonlinear ones:

$$\{\hat{A}, \hat{B}, \hat{C}_1, \hat{C}_2\} = \underset{A, B, C_1, C_2}{\arg\min}\, r^2(A, B, C_1, C_2, t_c, m, \omega), \quad \text{(S6)}$$

whose solution can be obtained by analytically solving the following system of linear equations:

$$\begin{bmatrix} n & \sum f_i & \sum g_i & \sum h_i \\ \sum f_i & \sum f_i^2 & \sum f_i g_i & \sum f_i h_i \\ \sum g_i & \sum f_i g_i & \sum g_i^2 & \sum g_i h_i \\ \sum h_i & \sum f_i h_i & \sum g_i h_i & \sum h_i^2 \end{bmatrix} \begin{bmatrix} \hat{A} \\ \hat{B} \\ \hat{C}_1 \\ \hat{C}_2 \end{bmatrix} = \begin{bmatrix} \sum \Omega_i \\ \sum \Omega_i f_i \\ \sum \Omega_i g_i \\ \sum \Omega_i h_i \end{bmatrix}, \quad \text{(S7)}$$

where $f_i = (t_c - t_i)^m$, $g_i = (t_c - t_i)^m \cos\left[\omega \ln(t_c - t_i)\right]$, and $h_i = (t_c - t_i)^m \sin\left[\omega \ln(t_c - t_i)\right]$. When solving the above calibration problem, we adopt a filter $4.94 \le \omega \le 15$, whose lower and upper bounds are defined to respectively prevent chaotic (Saleur et al., 1996b) and spurious (Filimonov and Sornette, 2013) scenarios.

**Text S2.** Global dataset of geohazard events

The dataset analyzed in this study comprises 109 historical geohazard events and builds on the compilation presented in our previous study (Lei and Sornette, 2025). Time series were obtained either directly from monitoring systems or digitized from published literature. Data sources and references are fully documented in Tables S1–S4.

**Text S3.** Statistical assessment of the scaling exponent

This section describes the statistical tests used to assess the significance of the scaling relation between precursory duration $T$ and failure volume $V$, expressed as $T \propto V^{\zeta}$. All analyses are performed using logarithmic variables:

$$x_j = \log_{10} V_j \text{ and } y_j = \log_{10} T_j, \tag{S8}$$

such that

$$y_j = \xi + \zeta x_j + \varepsilon_j \tag{S9}$$

where the index $j$ labels individual geohazard events in the dataset, $\xi$ is an intercept, and $\varepsilon_j$ denotes residual fluctuations.

The primary objective is to test whether a positive scaling exists between precursory duration and failure volume. Accordingly, the following one-sided hypotheses are evaluated for tests that explicitly estimate the scaling exponent $\zeta$:

$H_0$: $\zeta = 0$ (no scaling between precursory duration and failure volume);
$H_1$: $\zeta > 0$ (positive scaling between precursory duration and failure volume).

Statistical significance of the scaling exponent is assessed using multiple complementary parametric and non-parametric methods, as described below. All tests are applied both to the full dataset including all hazard types and independently to each hazard type (Table S5). This allows us to assess whether the observed scaling reflects a cross-hazard tendency or is driven by specific hazard classes. Failure to reject the null hypothesis for a given hazard type is interpreted as statistical non-detection of scaling, rather than evidence for its absence.

(1) Ordinary least squares regression and robust inference

The scaling exponent $\zeta$ is estimated by ordinary least squares. Because log–log data commonly exhibit heteroskedasticity, statistical inference is based on heteroskedasticity-consistent (White-type) robust standard errors. We report results using both the HC1 estimator, which includes a small-sample correction, and the more conservative HC3 estimator, which further accounts for leverage effects.

When robust standard errors are used, the finite-sample distribution of the $t$-statistic is not exactly Student-$t$; however, under standard regularity conditions, the statistic is asymptotically normally distributed. Accordingly, one-sided $p$-values are evaluated using both a conservative Student-$t$ approximation and the asymptotic normal approximation.

(2) Likelihood-ratio test

As a complementary parametric test, a likelihood-ratio test is performed by comparing the full model $y = \xi + \zeta x + \varepsilon$ with the nested null model $y = \xi + \varepsilon$. Under the assumption of Gaussian residuals, the likelihood-ratio statistic is asymptotically chi-squared distributed with one degree of freedom under the null hypothesis. It provides an independent assessment of whether inclusion of the scaling term significantly improves model fit.

(3) Permutation test

To minimize reliance on distributional assumptions, an assumption-light permutation test is conducted. The observed scaling exponent $\zeta$ is compared against a null distribution obtained by randomly permuting $y$ relative to $x$, thereby destroying any association while preserving the marginal distributions. For each permutation, the regression slope is re-estimated. The one-sided $p$-value is defined as the probability of obtaining a permuted slope greater than or equal to the observed $\zeta$.

(4) Bootstrap confidence intervals

Uncertainty in $\zeta$ is further quantified using a non-parametric case bootstrap, in which paired observations ($x_j$, $y_j$) are resampled with replacement. We report the one-sided 95% lower confidence bound. A lower bound exceeding zero provides non-parametric evidence against the null hypothesis $\zeta = 0$.

(5) Rank-based association tests

As complementary evidence, Pearson's correlation test, Spearman's rank correlation test, and Kendall's rank correlation test are performed. These tests do not directly evaluate the null hypothesis $\zeta = 0$, but instead assess whether a statistically significant linear or monotonic association exists between $x$ and $y$. The resulting $p$-values provide assumption-light support for the robustness of the observed scaling.

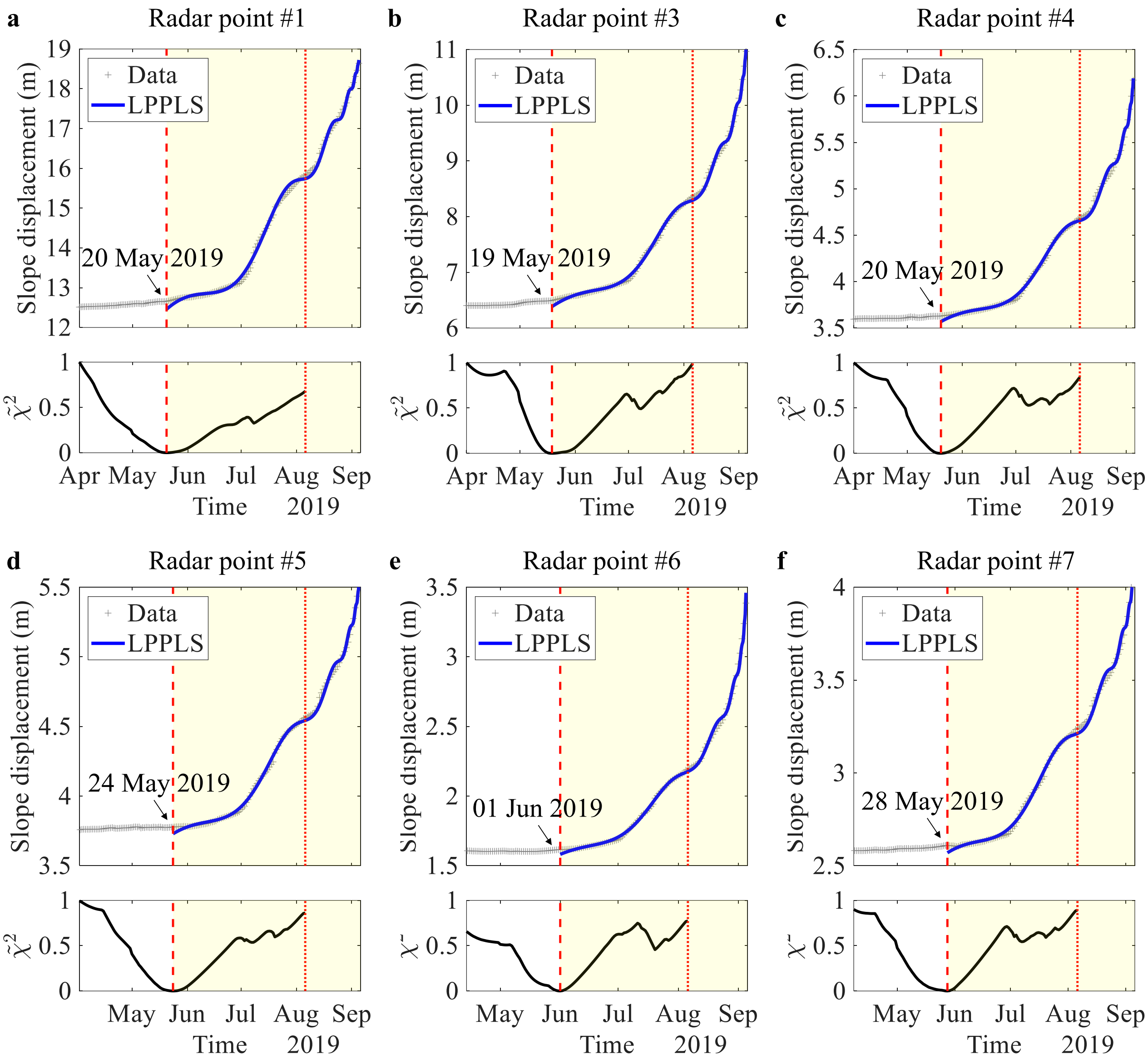

**Figure S1. Precursory dynamics and onset identification for the Veslemannen landslides based on different radar point measurements. In each subfigure, the upper panel shows the monitoring data (grey symbols) together with the corresponding LPPLS fit (blue line), while the lower panel shows the normalized regularized cost function $\tilde{\chi}^2$ as a function of the calibration window start time $\tau$. Vertical dashed lines indicate the identified onset time $\tau^*$ of the precursory phase, vertical dotted lines mark the latest time of the $\tau$-scanning interval, and shaded regions highlight the inferred precursory phase.**

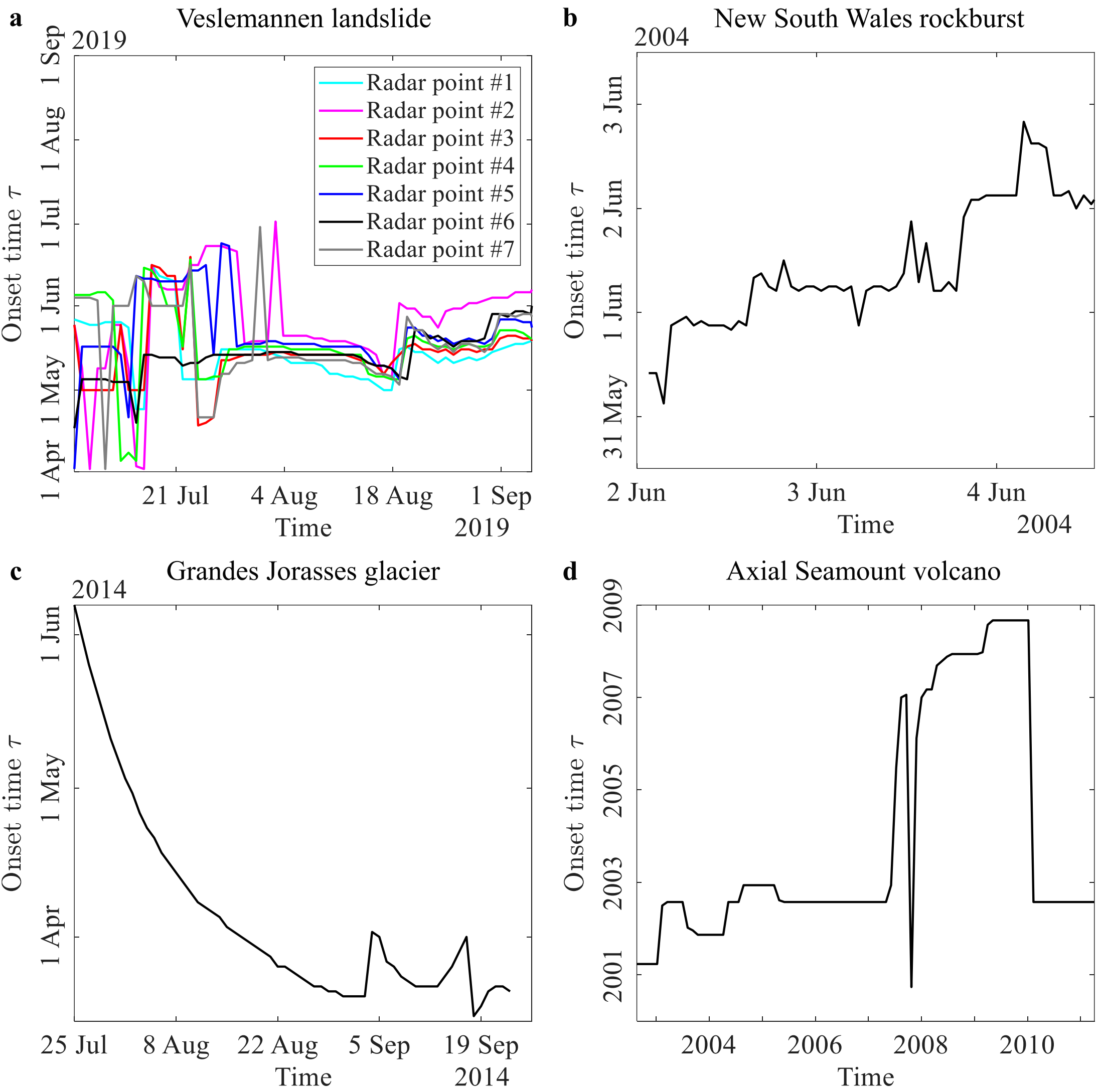

**Figure S2. Temporal evolution of the inferred onset time of precursory phase. The inferred onset time $\tau$ is shown as a function of the end time of the calibration window used for LPPLS fitting. Panels show examples for (a) the Veslemannen landslide, (b) the New South Wales rockburst, (c) the Grandes Jorasses glacier, and (d) the Axial Seamount volcano.**

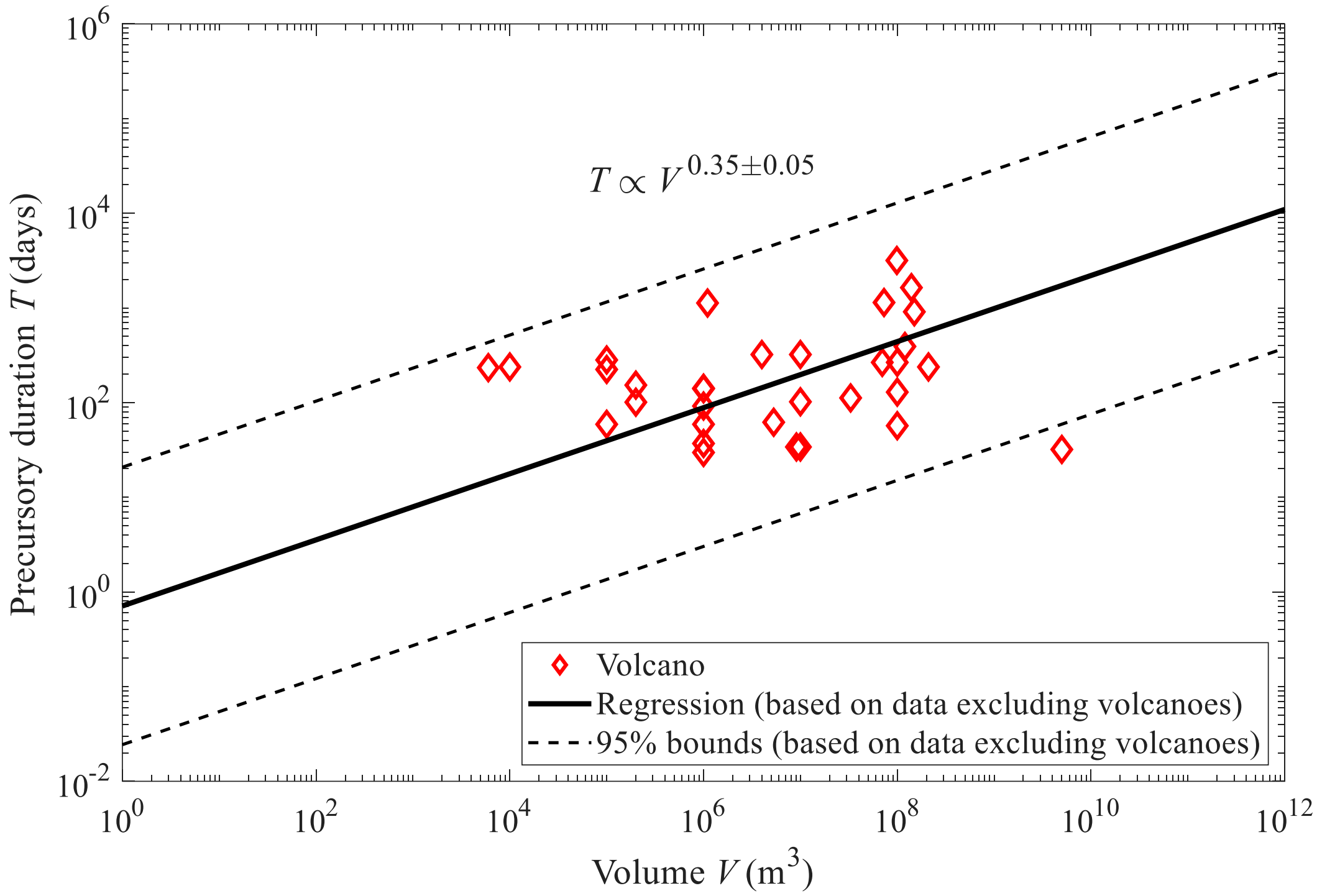

**Figure S3: Comparison of volcanic precursory durations with scaling derived from mechanically driven geohazards. The solid black line shows the scaling relationship between precursory duration and failure volume derived from the combined dataset of landslides, rockbursts, and glaciers, with dashed lines indicating the 95% confidence interval.**

**Table S1. Landslide cases (49 events in total).**

| Site | Location | Type | Material | Failure time | Volume ($m^3$) | Precursory duration (days) | Reference |
|---|---|---|---|---|---|---|---|
| Abbotsford | New Zealand | Soilslide | Clay | 1979-08-08 | $5×10^6$ | 19.4 | (Hancox, 2008) |
| Achoma | Peru | Rock-soilslide | Lacustrine sediments | 2020-06-16 | $5.4×10^6$ | 114.76 | (Lacroix et al., 2023) |
| Agoyama | Japan | Rockslide | Tuffaceous sandstone | 1972-12-02 | $\sim10^5$ | 42 | (Hayashi and Yamamori, 1991) |
| Arvigo | Switzerland | Topple / rockslide | Gneiss | 2007-05-28 | $2×10^5$ | 63 | (Leinauer et al., 2023) |
| Baishi | China | Rockslide | Phyllite | 2007-07-28 | $2×10^6$ | 31 | (Tang et al., 2024) |
| Baiyan | China | Rockslide | Limestone | 2022-05-08 | $2.5×10^4$ | 394.75 | (Li et al., 2023) |
| Brienz / Brinzauls | Switzerland | Rockslide | Flysch, schists, dolomite | 2023-06-15 | $1.2×10^6$ | 152 | (Loew et al., 2024) |
| Cadia | Australia | Soilslide | Earthfill materials | 2018-03-09 | $7.2×10^4$ | 298.86 | (Carlà et al., 2019a) |
| Copper open pit | Undisclosed | Rockslide | Limestone, spilite | 2016-11-17 | $6.4×10^5$ | 241.73 | (Carlà et al., 2019a) |
| Dosan | Japan | Rockslide | Schist | 1962-02-20 | $6×10^4$ | 2.17 | (Saito, 1965) |
| Gallivaggio | Italy | Rockfall | Granite | 2018-05-29 | $5×10^3$ | 26.18 | (Carlà et al., 2019b) |
| Galterengraben | Switzerland | Rockfall | Sandstone | 2016-04-24 | $2.5×10^3$ | 360 | (Leinauer et al., 2023) |
| Grabengufer | Switzerland | Rockfall | Rock & ice | 2020-05-17 | $5×10^2$ | 2.83 | (Cicoira et al., 2022) |
| Hogarth | Canada | Topple | Diorite | 1975-06-23 | $2×10^5$ | 80 | (Brawner and Stacey, 1979) |
| Iron mine | Mexico | Rockslide | Rock | 1990-07-26 | $\sim5×10^4$ | 39 | (Ryan and Call, 1992) |
| Jinlonggou | China | Rockslide | Syenite & basalt | 2010-10-23 | $2×10^5$ | 65 | (Chen et al., 2021) |
| Kagemori | Japan | Rockslide | Limestone | 1973-09-20 | $3×10^5$-$4×10^5$ | 140 | (Yamaguchi and Shimotani, 1986) |
| La Saxe | Italy | Rockslide | Meta-sedimentary sequences | 2013-04-21 | $5×10^2$-$1×10^3$ | 63 | (Manconi and Giordan, 2016) |
| Letlhakane diamond mine | Botswana | Rockslide | Sandstone | 2005-07-14 | $2.3×10^5$ | 97.22 | (Kayesa, 2006) |
| Longjing | China | Rockslide | Dolomite & limestone | 2019-02-17 | $1.4×10^6$ | 14.29 | (Fan et al., 2019) |
| Maoxian | China | Soilslide | Soil | 2017-06-24 | $1.5×10^7$ | 656.5 | (Intrieri et al., 2018) |
| Mt. Beni | Italy | Rockslide | Basalt & limestone | 2002-12-28 | $5.0×10^5$ | 233.23 | (Gigli et al., 2011) |
| Mud Greek | USA | Rockslide | Shale, sandstone, sediments | 2017-05-20 | $3×10^6$ | 472.8 | (Jacquemart and Tiampo, 2021) |
| Nevis Bluff | New Zealand | Topple / rockslide | Schist | 1975-06-14 | $3.2×10^4$ | 100 | (Brown et al., 1980) |
| New Tredegar | UK | Rockslide | Sandstone | 1930-04-12 | $\sim7×10^4$ | 37.94 | (Carey, 2011) |
| Northern Bohemia | Czech Republic | Rockfall | Sandstone | 1984-01-07 | $1.4×10^3$ | 36.58 | (Zvelebill and Moser, 2001) |

**Table S1 (continued). Landslide cases (49 events in total).**

| Site | Location | Type | Material | Failure time | Volume ($m^3$) | Precursory duration (days) | Reference |
|---|---|---|---|---|---|---|---|
| Open pit mine (event 3) | Undisclosed | Rockslide | Anorthosite | 2014-09-26 | $6\times10^2$ | 0.46 | (Carlà et al., 2017) |
| Open pit mine (event 4) | Undisclosed | Rockslide | Anorthosite | 2014-2017 | $3\times10^3$ | 1.4 | (Carlà et al., 2017) |
| Open pit mine (event 5) | Undisclosed | Topple | Anorthosite | 2017-02-05 | $4\times10^3$ | 2.79 | (Carlà et al., 2017) |
| Otomura | Japan | Rockslide | Sandstone & shale | 2004-08-10 | $2\times10^5$ | 48 | (Fujisawa et al., 2010) |
| Preonzo | Switzerland | Rockslide | Gneiss | 2012-05-15 | $2.1\times10^5$ | 66 | (Loew et al., 2017) |
| Puigcercós | Spain | Rockfall | Marl, silt, sandstone, limestone | 2013-12-03 | $1\times10^3$ | 2120.11 | (Royán et al., 2015) |
| Road slope (event 1) | Undisclosed | Rockslide | Mobilized gneiss | 2009-01-24 | $3\times10^2$ | 1.96 | (Mazzanti et al., 2015) |
| Road slope (event 2) | Undisclosed | Flow | Colluvium | 2009-02-18 | $1.4\times10^1$ | 9.33 | (Mazzanti et al., 2015) |
| Road slope (event 3) | Undisclosed | Soilslide | Colluvium & beton | 2009-12-20 | $1.6\times10^2$ | 0.17 | (Mazzanti et al., 2015) |
| Road slope (event 4) | Undisclosed | Soilslide | Colluvium & beton | 2010-01-16 | $2\times10^2$ | 0.08 | (Mazzanti et al., 2015) |
| Road slope (event 5) | Undisclosed | Soilslide | Colluvium & beton | 2009-02-03 | $8\times10^1$ | 0.26 | (Mazzanti et al., 2015) |
| Road slope (event 6) | Undisclosed | Soilslide | Colluvium & beton | 2010-02-11 | $5\times10^2$ | 1.58 | (Mazzanti et al., 2015) |
| Road slope (event 7) | Undisclosed | Flow | Mobilized & altered gneiss | 2010-02-12 | $2\times10^2$ | 0.83 | (Mazzanti et al., 2015) |
| Road slope (event 8) | Undisclosed | Rockslide | Colluvium & beton | 2010-02-12 | $3\times10^2$ | 0.21 | (Mazzanti et al., 2015) |
| Road slope (event 9) | Undisclosed | Rockslide | Mobilized & altered gneiss | 2010-02-17 | $8\times10^1$ | 9.92 | (Mazzanti et al., 2015) |
| Road slope (event 10) | Undisclosed | Rockslide | Mobilized & altered gneiss | 2010-03-10 | $1.5\times10^2$ | 0.68 | (Mazzanti et al., 2015) |
| Roesgrenda | Norway | Soilslide | Quick clay | 2000-03-02 | $2\times10^3$ | 35.55 | (Okamoto et al., 2004) |
| Takabayama | Japan | Rockslide | Mudstone, sandstone | 1970-01-22 | $5\times10^3$ | 46 | (Saito, 1979) |
| Vajont | Italy | Rockslide | Limestone | 1963-10-09 | $2.7\times10^8$ | 98 | (Nonveiller, 1987) |
| Veslemannen | Norway | Rockslide | Gneiss | 2019-09-05 | $5.4\times10^4$ | 109 | (Kristensen et al., 2021) |
| Welland | Canada | Soilslide | Clay | 1967-02-22 | $5\times10^2$ | 4.83 | (Kwan, 1971) |
| Xintan | China | Rockslide | Sediments | 1985-06-12 | $3\times10^7$ | 221.87 | (Xue et al., 2014) |
| Yusuihsi | Taiwan | Rockslide | Slate & phyllite | 2021-08-07 | $8.5\times10^6$ | 847 | (Kuo et al., 2025) |

**Table S2. Rockburst cases (11 events in total).**

| Site | Location | Material | Failure time | Volume ($m^3$) | Precursory duration (days) | Reference |
|---|---|---|---|---|---|---|
| Coal mine (cut-through #4) | Australia | Coal | 2004-06-11 | $\sim 5\times10^{1}$ | 1.33 | (Shen et al., 2008) |
| Coal mine (cut-through #5) | Australia | Coal | 2004-06-04 | $\sim 5\times10^{1}$ | 2.48 | (Shen et al., 2008) |
| Gold mine (event 1) | South Africa | Metamorphic rock | 1997-02-05 | $\sim 1\times10^{4}$ | 4.41 | (Ouillon and Sornette, 2000) |
| Gold mine (event 2) | South Africa | Metamorphic rock | 1997-04-03 | $\sim 5\times10^{3}$ | 1.34 | (Ouillon and Sornette, 2000) |
| Gold mine (event 3) | South Africa | Metamorphic rock | 1997-04-06 | $\sim 5\times10^{4}$ | 11.04 | (Ouillon and Sornette, 2000) |
| Gold mine (event 4) | South Africa | Metamorphic rock | 1997-05-04 | $\sim 3\times10^{4}$ | 43.01 | (Ouillon and Sornette, 2000) |
| Gold mine (event 5) | South Africa | Metamorphic rock | 1997-05-26 | $\sim 3\times10^{3}$ | 12.41 | (Ouillon and Sornette, 2000) |
| Gold mine (event 6) | South Africa | Metamorphic rock | 1997-06-06 | $\sim 1\times10^{4}$ | 5.77 | (Ouillon and Sornette, 2000) |
| Gold mine (event 7) | South Africa | Metamorphic rock | 1997-06-23 | $\sim 4\times10^{3}$ | 63.18 | (Ouillon and Sornette, 2000) |
| Gold mine (event 8) | South Africa | Metamorphic rock | 1997-08-15 | $\sim 4\times10^{3}$ | 114.2 | (Ouillon and Sornette, 2000) |
| Platinum mine | South Africa | Merensky reef | 2007-04-04 | $\sim 1\times10^{1}$ | 8.08 | (Malan et al., 2007) |

Note: the volumes of the rockburst events at the Gold mine are estimated based on the documented seismic moment.

**Table S3. Glacier cases (17 events in total).**

| Site | Location | Type | Failure time | Volume ($m^3$) | Precursory duration (days) | Reference |
|---|---|---|---|---|---|---|
| Amery | Antarctica | Ice shelf | 2019-09-25 | $3.4\times10^{11}$ | 2478 | (Walker et al., 2021) |
| Eiger glacier (2001 event) | Switzerland | Polythermal glacier | 2001-08-20 | $2.7\times10^{5}$ | 5.15 | (Pralong et al., 2005) |
| Eiger glacier (2016 event) | Switzerland | Polythermal glacier | 2016-08-25 | $1.5\times10^{4}$ | 39 | (Chmiel et al., 2023) |
| Grandes Jorasses (2014 event) | Italy | Cold glacier | 2014-09-23 & 2014-09-29 | $1.5\times10^{5}$ | 186 | (Faillettaz et al., 2016) |
| Grandes Jorasses (2020 event) | Italy | Cold glacier | 2020-11-11 | $2\times10^{4}$ | 161 | (Dematteis et al., 2021) |
| Gruben | Switzerland | Temperate glacier | 1974-09-09 | $8\times10^{3}$ | 3.21 | (Pralong et al., 2005) |
| Mönch | Switzerland | Temperate glacier | 2003-07-04 | $6\times10^{5}$ | 66 | (Pralong et al., 2005) |
| Planpincieux (event 1) | Italy | Polythermal glacier | 2015-08-16 | $1.4\times10^{4}$ | 67 | (Dematteis et al., 2021; Giordan et al., 2020) |
| Planpincieux (event 2) | Italy | Polythermal glacier | 2016-08-15 | $2.5\times10^{4}$ | 57 | (Dematteis et al., 2021; Giordan et al., 2020) |
| Planpincieux (event 3) | Italy | Polythermal glacier | 2017-08-03 | $2\times10^{4}$ | 52 | (Dematteis et al., 2021; Giordan et al., 2020) |
| Planpincieux (event 4) | Italy | Polythermal glacier | 2017-08-31 | $5.4\times10^{4}$ | 20 | (Dematteis et al., 2021; Giordan et al., 2020) |
| Planpincieux (event 5) | Italy | Polythermal glacier | 2017-10-12 | $1.5\times10^{4}$ | 30 | (Dematteis et al., 2021; Giordan et al., 2020) |
| Planpincieux (event 6) | Italy | Polythermal glacier | 2019-07-26 | $2.2\times10^{4}$ | 46 | (Dematteis et al., 2021; Giordan et al., 2020) |
| UK211 | Antarctica | Iceberg | 2006-11-23 | $2.4\times10^{9}$ | 428.43 | (Scambos et al., 2008) |
| Weisshorn (1973 event) | Switzerland | Cold glacier | 1972-10-17 | $5\times10^{5}$ | 252.7 | (Faillettaz et al., 2008) |
| Weisshorn (2005 event) | Switzerland | Cold glacier | 2005-03-30 | $5\times10^{5}$ | 23.26 | (Faillettaz et al., 2008, 2011) |
| Weissmies | Switzerland | Polythermal glacier | 2017-09-10 | $2.5\times10^{5}$ | 75 | (Leinauer et al., 2023; Meier et al., 2018) |

**Table S4. Volcano cases (32 events in total).**

| Site | Location | Type | Eruption time | Volume ($m^3$) | Precursory duration (days) | Reference |
|---|---|---|---|---|---|---|
| Adatara | Japan | Stratovolcano | 1996-09-01 | $1.1\times10^6$ | 1127 | (Newhall et al., 2017) |
| Asama | Japan | Complex volcano | 2009-02-02 | $\sim10^4$ | 238 | (Newhall et al., 2017) |
| Augustine | USA | Lava dome | 2006-01-11 | $7.3\times10^7$ | 1141 | (Newhall et al., 2017) |
| Axial Seamount | Pacific Ocean | Submarine fissure volcano | 2011-04-06 | $9.9\times10^7$ | 3171 | (Chadwick et al., 2022) |
| Bezymianny | Russia | Stratovolcano | 1960-04-10 | $\sim10^6$ | 29.72 | (Voight, 1988) |
| Etna (1989 event) | Italy | Stratovolcano | 1989-09-08 | $\sim10^7$ | 322 | (Newhall et al., 2017) |
| Etna (2013 event) | Italy | Stratovolcano | 2013-09-05 | $\sim10^6$ | 140 | (Cannata et al., 2015; Newhall et al., 2017) |
| Hierro | Spain | Submarine shield volcano | 2011-10-10 | $3.3\times10^7$ | 112 | (Carracedo et al., 2015; Newhall et al., 2017) |
| Kilauea (1971 event) | USA | Shield volcano | 1971-08-14 | $9.1\times10^6$ | 34 | (Newhall et al., 2017) |
| Kilauea (1972 event) | USA | Shield volcano | 1972-02-04 | $1.2\times10^8$ | 392 | (Newhall et al., 2017) |
| Kilauea (1983 event) | USA | Shield volcano | 1983-01-03 | $4\times10^6$ | 322 | (Newhall et al., 2017) |
| Kujusan | Japan | Stratovolcano | 1995-10-11 | $2\times10^5$ | 101 | (Newhall et al., 2017; Sudo et al., 1998) |
| Mauna Loa | USA | Shield volcano | 1984-03-25 | $\sim10^8$ | 129 | (Bell et al., 2011) |
| Merapi (2006 event) | Indonesia | Stratovolcano | 2006-06-06 | $5.3\times10^6$ | 62 | (Newhall et al., 2017; Ratdomopurbo et al., 2013) |
| Merapi (2010 event) | Indonesia | Stratovolcano | 2010-10-26 | $\sim10^7$ | 34 | (Newhall et al., 2017; Surono et al., 2012) |
| Pinatubo | Philippines | Stratovolcano | 1991-06-07 | $5\times10^9$ | 32 | (Newhall et al., 2017) |
| Redoubt (1989 event) | USA | Stratovolcano | 1989-12-14 | $\sim10^8$ | 266 | (Miller and Chouet, 1994; Newhall et al., 2017) |
| Redoubt (2009 event) | USA | Stratovolcano | 2009-03-22 | $\sim10^8$ | 57 | (Bull and Buurman, 2013; Newhall et al., 2017) |
| Ruapehu (1995 event) | New Zealand | Stratovolcano | 1995-09-18 | $2\times10^5$ | 153 | (Newhall et al., 2017) |
| Ruapehu (1996 event) | New Zealand | Stratovolcano | 1996-06-19 | $\sim10^7$ | 102 | (Newhall et al., 2017) |
| Ruapehu (2006 event) | New Zealand | Stratovolcano | 2006-10-04 | $\sim10^5$ | 282 | (Newhall et al., 2017) |
| Sakurajima | Japan | Stratovolcano | 2017-03-25 | $\sim10^6$ | 92 | (Newhall et al., 2017) |
| Sierra Negra (2005 event) | Ecuador | Shield volcano | 2005-10-22 | $1.5\times10^8$ | 910 | (Geist et al., 2008) |
| Sierra Negra (2018 event) | Ecuador | Shield volcano | 2018-06-26 | $1.4\times10^8$ | 1638 | (Bell et al., 2021) |

**Table S4 (continued). Volcano cases (32 events in total).**

| Site | Location | Type | Eruption time | Volume ($m^3$) | Precursory duration (days) | Reference |
|---|---|---|---|---|---|---|
| St. Helens (1980 event) | USA | Stratovolcano | 1980-03-27 | $\sim10^5$ | 59 | (Newhall et al., 2017; Swanson et al., 1983) |
| St. Helens (1981 event) | USA | Stratovolcano | 1981-09-06 | $\sim10^6$ | 142 | (Newhall et al., 2017; Swanson et al., 1983) |
| St. Helens (1982 event) | USA | Stratovolcano | 1982-03-19 | $\sim10^6$ | 59 | (Newhall et al., 2017; Swanson et al., 1983) |
| St. Helens (1985 event) | USA | Stratovolcano | 1985-05-25 | $\sim10^6$ | 37 | (Bell et al., 2013; Newhall et al., 2017) |
| Soufriere Hills | UK | Stratovolcano | 1995-11-15 | $7\times10^7$ | 265.86 | (Newhall et al., 2017; Young et al., 1998) |
| Tokachidake | Japan | Stratovolcano | 2004-02-25 | $\sim10^5$ | 224 | (Newhall et al., 2017) |
| Unzen | Japan | Complex volcano | 1990-11-17 | $2.1\times10^8$ | 238 | (Nakada et al., 1999; Newhall et al., 2017) |
| Yakedake | Japan | Stratovolcano | 1995-02-11 | $6\times10^3$ | 234 | (Newhall et al., 2017) |

**Table S5. Statistical tests of the scaling exponent $\zeta$.**

| Hazard type | Sample size | $\zeta$ | $P$ (HC1) | $p$ (HC3) | $p$ (LRT) | $p$ (Permutation) | Bootstrap 95% lower bound | $p$ (Pearson) | $p$ (Spearman) | $p$ (Kendall) |
|---|---|---|---|---|---|---|---|---|---|---|
| All (incl. volcanoes) | 109 | 0.314 | $< 10^{-12}$ | $< 10^{-11}$ | $< 10^{-14}$ | 0 | 0.251 | $< 10^{-14}$ | $< 10^{-12}$ | $< 10^{-11}$ |
| All (excl. volcanoes) | 77 | 0.35 | $< 10^{-10}$ | $< 10^{-10}$ | $< 10^{-10}$ | 0 | 0.281 | $< 10^{-10}$ | $< 10^{-10}$ | $< 10^{-9}$ |
| Landslides | 49 | 0.40 | $< 10^{-6}$ | $< 10^{-6}$ | $< 10^{-6}$ | 0 | 0.295 | $< 10^{-6}$ | $< 10^{-7}$ | $< 10^{-6}$ |
| Rockbursts | 11 | 0.21 | 0.053 | 0.10 | 0.17 | 0.115 | 0.018 | 0.228 | 0.498 | 0.480 |
| Glaciers | 17 | 0.24 | $< 10^{-5}$ | $< 10^{-4}$ | $< 10^{-3}$ | 0.001 | 0.081 | 0.001 | 0.051 | 0.039 |
| Volcanoes | 32 | 0.061 | 0.20 | 0.22 | 0.36 | 0.188 | –0.040 | 0.376 | 0.225 | 0.312 |

Note: Results are shown for the full dataset (including or excluding volcanoes, while including all other types) and for each hazard type analyzed independently. The scaling exponent $\zeta$ is estimated by ordinary least squares regression of $\log_{10}T$ on $\log_{10}V$. Statistical significance is assessed using heteroskedasticity-consistent robust standard errors (HC1 and HC3) with one-sided tests of $H_0$: $\zeta = 0$ versus $H_1$: $\zeta > 0$, a likelihood-ratio test (LRT), a permutation test, and bootstrap confidence intervals. Pearson and rank-based (Spearman and Kendall) correlation tests are reported as complementary evidence for linear and monotonic association, respectively. In the table, $p$-values are one-sided unless stated otherwise; the bootstrap lower bound refers to the one-sided 95% confidence bound; a negative lower bound indicates that $\zeta = 0$ cannot be rejected.

## References

Bell, A. F., Greenhough, J., Heap, M. J., and Main, I. G.: Challenges for forecasting based on accelerating rates of earthquakes at volcanoes and laboratory analogues, Geophys. J. Int., 185, 718–723, https://doi.org/10.1111/j.1365-246X.2011.04982.x, 2011.

Bell, A. F., Naylor, M., and Main, I. G.: The limits of predictability of volcanic eruptions from accelerating rates of earthquakes, Geophys. J. Int., 194, 1541–1553, https://doi.org/10.1093/gji/ggt191, 2013.

Bell, A. F., La Femina, P. C., Ruiz, M., Amelung, F., Bagnardi, M., Bean, C. J., Bernard, B., Ebinger, C., Gleeson, M., Grannell, J., Hernandez, S., Higgins, M., Liorzou, C., Lundgren, P., Meier, N. J., Möllhoff, M., Oliva, S.-J., Ruiz, A. G., and Stock, M. J.: Caldera resurgence during the 2018 eruption of Sierra Negra volcano, Galápagos Islands, Nat. Commun., 12, 1397, https://doi.org/10.1038/s41467-021-21596-4, 2021.

Brawner, C. O. and Stacey, P. F.: Hogarth Pit Slope Failure, Ontario, Canada, in: Developments in Geotechnical Engineering, vol. 14, edited by: Voight, B., Elsevier, 691–707, 1979.

Brown, I., Hittinger, M., and Goodman, R.: Finite element study of the Nevis Bluff (New Zealand) rock slope failure, Rock Mech., 12, 231–245, https://doi.org/10.1007/BF01251027, 1980.

Bull, K. F. and Buurman, H.: An overview of the 2009 eruption of Redoubt Volcano, Alaska, J. Volcanol. Geotherm. Res., 259, 2–15, https://doi.org/10.1016/j.jvolgeores.2012.06.024, 2013.

Cannata, A., Spedalieri, G., Behncke, B., Cannavò, F., Di Grazia, G., Gambino, S., Gresta, S., Gurrieri, S., Liuzzo, M., and Palano, M.: Pressurization and depressurization phases inside the plumbing system of Mount Etna volcano: Evidence from a multiparametric approach, J. Geophys. Res. Solid Earth, 120, 5965–5982, https://doi.org/10.1002/2015JB012227, 2015.

Carey, J.: The Progressive Development and Post-failure Behaviour of Deep-seated Landslide Complexes, Doctoral, Durham University, 2011.

Carlà, T., Farina, P., Intrieri, E., Botsialas, K., and Casagli, N.: On the monitoring and early-warning of brittle slope failures in hard rock masses: Examples from an open-pit mine, Engineering Geology, 228, 71–81, https://doi.org/10.1016/j.enggeo.2017.08.007, 2017.

Carlà, T., Intrieri, E., Raspini, F., Bardi, F., Farina, P., Ferretti, A., Colombo, D., Novali, F., and Casagli, N.: Perspectives on the prediction of catastrophic slope failures from satellite InSAR, Sci. Rep., 9, 14137, https://doi.org/10.1038/s41598-019-50792-y, 2019a.

Carlà, T., Nolesini, T., Solari, L., Rivolta, C., Dei Cas, L., and Casagli, N.: Rockfall forecasting and risk management along a major transportation corridor in the Alps through ground-based radar interferometry, Landslides, 16, 1425–1435, https://doi.org/10.1007/s10346-019-01190-y, 2019b.

Carracedo, J. C., Troll, V. R., Zaczek, K., Rodríguez-González, A., Soler, V., and Deegan, F. M.: The 2011–2012 submarine eruption off El Hierro, Canary Islands: New lessons in oceanic island growth and volcanic crisis management, Earth-Sci. Rev., 150, 168–200, https://doi.org/10.1016/j.earscirev.2015.06.007, 2015.

Chadwick, W. W., Wilcock, W. S. D., Nooner, S. L., Beeson, J. W., Sawyer, A. M., and Lau, T. -K.: Geodetic monitoring at Axial Seamount since its 2015 eruption reveals a waning magma supply and tightly linked rates of deformation and seismicity, Geochem. Geophys. Geosyst., 23, e2021GC010153, https://doi.org/10.1029/2021GC010153, 2022.

Chen, M., Huang, D., and Jiang, Q.: Slope movement classification and new insights into failure prediction based on landslide deformation evolution, Int. J. Rock Mech. Min. Sci., 141, 104733, https://doi.org/10.1016/j.ijrmms.2021.104733, 2021.

Chmiel, M., Walter, F., Pralong, A., Preiswerk, L., Funk, M., Meier, L., and Brenguier, F.: Seismic constraints on damage growth within an unstable hanging glacier, Geophys. Res. Lett., 50, e2022GL102007, https://doi.org/10.1029/2022GL102007, 2023.

Cicoira, A., Weber, S., Biri, A., Buchli, B., Delaloye, R., Da Forno, R., Gärtner-Roer, I., Gruber, S., Gsell, T., Hasler, A., Lim, R., Limpach, P., Mayoraz, R., Meyer, M., Noetzli, J., Phillips, M., Pointner, E., Raetzo, H., Scapozza, C., Strozzi, T., Thiele, L., Vieli, A., Vonder Mühll, D., Wirz, V., and Beutel, J.: In situ observations of the Swiss periglacial environment using GNSS instruments, Earth Syst. Sci. Data, 14, 5061–5091, https://doi.org/10.5194/essd-14-5061-2022, 2022.

Dematteis, N., Giordan, D., Troilo, F., Wrzesniak, A., and Godone, D.: Ten-year monitoring of the Grandes Jorasses glaciers kinematics. Limits, potentialities, and possible applications of different monitoring systems, Remote Sens., 13, 3005, https://doi.org/10.3390/rs13153005, 2021.

Faillettaz, J., Pralong, A., Funk, M., and Deichmann, N.: Evidence of log-periodic oscillations and increasing icequake activity during the breaking-off of large ice masses, J. Glaciol., 54, 725–737, https://doi.org/10.3189/002214308786570845, 2008.

Faillettaz, J., Funk, M., and Sornette, D.: Icequakes coupled with surface displacements for predicting glacier break-off, J. Glaciol., 57, 453–460, https://doi.org/10.3189/002214311796905668, 2011.

Faillettaz, J., Funk, M., and Vagliasindi, M.: Time forecast of a break-off event from a hanging glacier, Cryosphere, 10, 1191–1200, https://doi.org/10.5194/tc-10-1191-2016, 2016.

Fan, X., Xu, Q., Liu, J., Subramanian, S. S., He, C., Zhu, X., and Zhou, L.: Successful early warning and emergency response of a disastrous rockslide in Guizhou province, China, Landslides, 16, 2445–2457, https://doi.org/10.1007/s10346-019-01269-6, 2019.

Filimonov, V. and Sornette, D.: A stable and robust calibration scheme of the log-periodic power law model, Physica A Stat. Mech. Appl., 392, 3698–3707, https://doi.org/10.1016/j.physa.2013.04.012, 2013.

Fujisawa, K., Marcato, G., Nomura, Y., and Pasuto, A.: Management of a typhoon-induced landslide in Otomura (Japan), Geomorphology, 124, 150–156, https://doi.org/10.1016/j.geomorph.2010.09.027, 2010.

Geist, D. J., Harpp, K. S., Naumann, T. R., Poland, M., Chadwick, W. W., Hall, M., and Rader, E.: The 2005 eruption of Sierra Negra volcano, Galápagos, Ecuador, Bull. Volcanol., 70, 655–673, https://doi.org/10.1007/s00445-007-0160-3, 2008.

Gigli, G., Fanti, R., Canuti, P., and Casagli, N.: Integration of advanced monitoring and numerical modeling techniques for the complete risk scenario analysis of rockslides: The case of Mt. Beni (Florence, Italy), Eng. Geol., 120, 48–59, https://doi.org/10.1016/j.enggeo.2011.03.017, 2011.

Giordan, D., Dematteis, N., Allasia, P., and Motta, E.: Classification and kinematics of the Planpincieux Glacier break-offs using photographic time-lapse analysis, J. Glaciol., 66, 188–202, https://doi.org/10.1017/jog.2019.99, 2020.

Hancox, G. T.: The 1979 Abbotsford Landslide, Dunedin, New Zealand: a retrospective look at its nature and causes, Landslides, 5, 177–188, https://doi.org/10.1007/s10346-007-0097-9, 2008.

Hayashi, S. and Yamamori, T.: Forecast of time-to-slope failure by the a-tr method, J. Japan Landslide Soc., 28, 1–8, https://doi.org/10.3313/jls1964.28.2_1, 1991.

Intrieri, E., Raspini, F., Fumagalli, A., Lu, P., Del Conte, S., Farina, P., Allievi, J., Ferretti, A., and Casagli, N.: The Maoxian landslide as seen from space: detecting precursors of failure with Sentinel-1 data, Landslides, 15, 123–133, https://doi.org/10.1007/s10346-017-0915-7, 2018.

Jacquemart, M. and Tiampo, K.: Leveraging time series analysis of radar coherence and normalized difference vegetation index ratios to characterize pre-failure activity of the Mud Creek landslide, California, Nat. Hazards Earth Syst. Sci., 21, 629–642, https://doi.org/10.5194/nhess-21-629-2021, 2021.

Kayesa, G.: Prediction of slope failure at Letlhakane Mine with the Geomos slope monitoring system, in: Proceedings of the International Symposium on Stability of Rock Slopes in Open Pit Mining and Civil Engineering, International Symposium on Stability of Rock Slopes in Open Pit Mining and Civil Engineering, 2006.

Kristensen, L., Czekirda, J., Penna, I., Etzelmüller, B., Nicolet, P., Pullarello, J. S., Blikra, L. H., Skrede, I., Oldani, S., and Abellan, A.: Movements, failure and climatic control of the Veslemannen rockslide, Western Norway, Landslides, 18, 1963–1980, https://doi.org/10.1007/s10346-020-01609-x, 2021.

Kuo, H.-L., Lin, G.-W., Lin, T.-Y., Liu, C.-H., Chu, C.-R., Chang, C.-H., Lin, C.-W., and Chen, H.: Displacement evolution of failure and non-failure sliding rock slopes, Landslides, 22, 1213–1226, https://doi.org/10.1007/s10346-024-02415-5, 2025.

Kwan, D.: Observations of the failure of a vertical cut in clay at Welland, Ontario, Can. Geotech. J., 8, 283–298, https://doi.org/10.1139/t71-024, 1971.

Lacroix, P., Huanca, J., Albinez, L., and Taipe, E.: Precursory motion and time-of-failure prediction of the Achoma landslide, Peru, from high frequency PlanetScope satellites, Geophys. Res. Lett., 50, e2023GL105413, https://doi.org/10.1029/2023GL105413, 2023.

Lei, Q. and Sornette, D.: Unified failure model for landslides, rockbursts, glaciers, and volcanoes., Commun. Earth Environ, 6, 390, https://doi.org/10.1038/s43247-025-02369-z, 2025.

Leinauer, J., Weber, S., Cicoira, A., Beutel, J., and Krautblatter, M.: An approach for prospective forecasting of rock slope failure time, Commun. Earth Environ., 4, 253, https://doi.org/10.1038/s43247-023-00909-z, 2023.

Li, B., Zhao, C., Li, J., Chen, H., Gao, Y., Cui, F., and Wan, J.: Mechanism of mining-induced landslides in the karst mountains of Southwestern China: a case study of the Baiyan landslide in Guizhou, Landslides, 20, 1481–1495, https://doi.org/10.1007/s10346-023-02047-1, 2023.

Loew, S., Gschwind, S., Gischig, V., Keller-Signer, A., and Valenti, G.: Monitoring and early warning of the 2012 Preonzo catastrophic rockslope failure, Landslides, 14, 141–154, https://doi.org/10.1007/s10346-016-0701-y, 2017.

Loew, S., Schneider, S., Josuran, M., Figi, D., Thoeny, R., Huwiler, A., Largiadèr, A., and Naenni, C.: Early warning and dynamics of compound rockslides: lessons learnt from the Brienz/Brinzauls 2023 rockslope failure, Landslides, https://doi.org/10.1007/s10346-024-02380-z, 2024.

Malan, D. F., Napier, J. A. L., and Janse van Rensburg, A. L.: Stope deformation measurements as a diagnostic measure of rock behaviour: A decade of research, J. South. Afr. Inst. Min. Metall., 107, 743–765, 2007.

Manconi, A. and Giordan, D.: Landslide failure forecast in near-real-time, Geomat. Nat. Haz. Risk, 7, 639–648, https://doi.org/10.1080/19475705.2014.942388, 2016.

Mazzanti, P., Bozzano, F., Cipriani, I., and Prestininzi, A.: New insights into the temporal prediction of landslides by a terrestrial SAR interferometry monitoring case study, Landslides, 12, 55–68, https://doi.org/10.1007/s10346-014-0469-x, 2015.

Meier, L., Jacquemart, M., Steinacher, R., Jäger, D., and Funk, M.: Monitoring of the Weissmies glacier before the failure event of September 10, 2017 with Radar Interferometry and high-resolution deformation camera, in: Proceedings of the International Snow Science Workshop, International Snow Science Workshop, 2018.

Miller, T. P. and Chouet, B. A.: The 1989–1990 eruptions of Redoubt Volcano: An introduction, J. Volcanol. Geotherm. Res., 62, 1–10, https://doi.org/10.1016/0377-0273(94)90025-6, 1994.

Nakada, S., Shimizu, H., and Ohta, K.: Overview of the 1990–1995 eruption at Unzen Volcano, J. Volcanol. Geotherm. Res., 89, 1–22, https://doi.org/10.1016/S0377-0273(98)00118-8, 1999.

Newhall, C. G., Costa, F., Ratdomopurbo, A., Venezky, D. Y., Widiwijayanti, C., Win, N. T. Z., Tan, K., and Fajiculay, E.: WOVOdat – An online, growing library of worldwide volcanic unrest, J. Volcanol. Geotherm. Res., 345, 184–199, https://doi.org/10.1016/j.jvolgeores.2017.08.003, 2017.

Nonveiller, E.: The Vajont reservoir slope failure, Eng. Geol., 24, 493–512, https://doi.org/10.1016/0013-7952(87)90081-0, 1987.

Okamoto, T., Larsen, J. O., Matsuura, S., Asano, S., Takeuchi, Y., and Grande, L.: Displacement properties of landslide masses at the initiation of failure in quick clay deposits and the effects of meteorological and hydrological factors, Eng. Geol., 72, 233–251, https://doi.org/10.1016/j.enggeo.2003.09.004, 2004.

Ouillon, G. and Sornette, D.: The concept of "critical earthquakes" applied to mine rockbursts with time-to-failure analysis, Geophys. J. Int., 143, 454–468, https://doi.org/10.1046/j.1365-246X.2000.01257.x, 2000.

Pralong, A., Birrer, C., Stahel, W. A., and Funk, M.: On the predictability of ice avalanches, Nonlin. Processes Geophys., 12, 849–861, https://doi.org/10.5194/npg-12-849-2005, 2005.

Ratdomopurbo, A., Beauducel, F., Subandriyo, J., Agung Nandaka, I. G. M., Newhall, C. G., Suharna, Sayudi, D. S., Suparwaka, H., and Sunarta: Overview of the 2006 eruption of Mt. Merapi, Journal of Volcanology and Geothermal Research, 261, 87–97, https://doi.org/10.1016/j.jvolgeores.2013.03.019, 2013.

Royán, M. J., Abellán, A., and Vilaplana, J. M.: Progressive failure leading to the 3 December 2013 rockfall at Puigcercós scarp (Catalonia, Spain), Landslides, 12, 585–595, https://doi.org/10.1007/s10346-015-0573-6, 2015.

Ryan, T. M. and Call, R. D.: Applications of rock mass monitoring for stability assessment of pit slope failure, in: Proceedings of the 33rd U.S. Symposium on Rock Mechanics, The 33rd U.S. Symposium on Rock Mechanics, 1992.

Saito, M.: Forecasting the time of occurrence of a slope failure, in: Proceedings of the 6th International Conference of Soil Mechanics and Foundation Engineering, 537–541, 1965.

Saito, M.: Evidencial study on forecasting occurence of slope failure, Japan, 1979.

Saleur, H., Sammis, C. G., and Sornette, D.: Renormalization group theory of earthquakes, Nonlinear Process. Geophys., 3, 102–109, https://doi.org/10.5194/npg-3-102-1996, 1996.

Scambos, T., Ross, R., Bauer, R., Yermolin, Y., Skvarca, P., Long, D., Bohlander, J., and Haran, T.: Calving and ice-shelf break-up processes investigated by proxy: Antarctic tabular iceberg evolution during northward drift, J. Glaciol., 54, 579–591, https://doi.org/10.3189/002214308786570836, 2008.

Shen, B., King, A., and Guo, H.: Displacement, stress and seismicity in roadway roofs during mining-induced failure, Int. J. Rock Mech. Min. Sci., 45, 672–688, https://doi.org/10.1016/j.ijrmms.2007.08.011, 2008.

Sudo, Y., Ono, H., Hurst, A. W., Tsutsui, T., Mori, T., Nakaboh, M., Matsumoto, Y., Sako, M., Yoshikawa, S., Tanaka, M., Kobayashi, Y., Hashimoto, T., Hoka, T., Yamada, T., Masuda, H., and Kikuchi, S.: Seismic activity and ground deformation associated with 1995 phreatic eruption of Kuju Volcano, Kyushu, Japan, J. Volcanol. Geotherm. Res., 81, 245–267, https://doi.org/10.1016/S0377-0273(98)00011-0, 1998.

Surono, Jousset, P., Pallister, J., Boichu, M., Buongiorno, M. F., Budisantoso, A., Costa, F., Andreastuti, S., Prata, F., Schneider, D., Clarisse, L., Humaida, H., Sumarti, S., Bignami, C., Griswold, J., Carn, S., Oppenheimer, C., and Lavigne, F.: The 2010 explosive eruption of Java's Merapi volcano—A '100-year' event, J. Volcanol. Geotherm. Res., 241–242, 121–135, https://doi.org/10.1016/j.jvolgeores.2012.06.018, 2012.

Swanson, D. A., Casadevall, T. J., Dzurisin, D., Malone, S. D., Newhall, C. G., and Weaver, C. S.: Predicting eruptions at Mount St. Helens, June 1980 through December 1982, Science, 221, 1369–1376, https://doi.org/10.1126/science.221.4618.1369, 1983.

Tang, R., Ren, S., Li, J., Feng, P., Li, H., Deng, R., Li, D., and Kasama, K.: The failure mechanism of the Baishi landslide in Beichuan County, Sichuan, China, Sci. Rep., 14, 17482, https://doi.org/10.1038/s41598-024-67402-1, 2024.
Voight, B.: A method for prediction of volcanic eruptions, Nature, 332, 125–130, https://doi.org/10.1038/332125a0, 1988.
Walker, C. C., Becker, M. K., and Fricker, H. A.: A high resolution, three-dimensional view of the D-28 Calving Event from Amery Ice Shelf with ICESat-2 and satellite imagery, Geophys. Res. Lett., 48, e2020GL091200, https://doi.org/10.1029/2020GL091200, 2021.
Xue, L., Qin, S., Li, P., Li, G., Adewuyi Oyediran, I., and Pan, X.: New quantitative displacement criteria for slope deformation process: From the onset of the accelerating creep to brittle rupture and final failure, Eng. Geol., 182, 79–87, https://doi.org/10.1016/j.enggeo.2014.08.007, 2014.
Yamaguchi, U. and Shimotani, T.: A case study of slope failure in a limestone quarry, Int. J. Rock Mech. Min. Sci. Geomech. Abstr., 23, 95–104, https://doi.org/10.1016/0148-9062(86)91670-0, 1986.
Young, S. R., Sparks, R. S. J., Aspinall, W. P., Lynch, L. L., Miller, A. D., Robertson, R. E. A., and Shepherd, J. B.: Overview of the eruption of Soufriere Hills Volcano, Montserrat, 18 July 1995 to December 1997, Geophys. Res. Lett., 25, 3389–3392, https://doi.org/10.1029/98GL01405, 1998.
Zvelebill, J. and Moser, M.: Monitoring based time-prediction of rock falls: Three case-histories, Phys. Chem. Earth Part B, 26, 159–167, https://doi.org/10.1016/S1464-1909(00)00234-3, 2001.